\documentclass[review]{elsarticle}
\usepackage{}

\usepackage[latin1]{inputenc} 
\usepackage{listings, tcolorbox}

\usepackage{lineno,hyperref}
\usepackage{amssymb}
\usepackage{color}
\usepackage{amsmath}
\usepackage{tikz}
\usepackage{amsfonts}
\usepackage{mathrsfs}
\usepackage{amsthm}

\usepackage{bbding}
\usepackage{mathrsfs}
\usepackage{amsmath, amsfonts, amssymb, mathrsfs, txfonts}
\usepackage{graphicx, subfigure}
\usepackage{wasysym}
\usepackage{color}
\usepackage{bm}
\usepackage{enumerate}

\usepackage{pifont}
\usepackage{txfonts}
\usepackage{amsmath}
\usepackage{graphicx}
\usepackage{amsfonts}
\usepackage{amssymb}
\usepackage{mathrsfs,psfrag,eepic,epsfig}
\usepackage{epstopdf}

\modulolinenumbers[5]

\journal{Journal of \LaTeX\ Templates}

\makeatletter \@addtoreset{equation}{section}

\newtheorem{thm}{Theorem}[section]

\newtheorem{lem}[thm]{Lemma}
\newtheorem{prop}[thm]{Proposition}
\theoremstyle{definition}

\begin{document}

\begin{frontmatter}

\title{Inverse scattering transform and multiple high-order pole solutions for the nonlocal focusing and defocusing modified Korteweg-de Vries  equation with the nonzero boundary conditions}
\tnotetext[mytitlenote]{
Corresponding author.\\
\hspace*{3ex}\emph{E-mail addresses}: sftian@cumt.edu.cn,
shoufu2006@126.com (S. F. Tian) }

\author{Xiao-Fan Zhang, Shou-Fu Tian$^{*}$ and Jin-Jie Yang}
\address{School of Mathematics, China University of Mining and Technology,  Xuzhou 221116, People's Republic of China}

\begin{abstract}
We extend the Riemann-Hilbert (RH) method to study the inverse scattering transformation and high-order pole solutions of the focusing and defocusing nonlocal (reverse-space-time) modified Korteweg-de Vries (mKdV) equations with nonzero boundary conditions (NZBCs) at infinity and successfully find its multiple soliton solutions with one high-order pole and multiple high-order poles. By introducing the generalized residue formula, we overcome the difficulty caused by  calculating the residue conditions corresponding to the higher-order poles.
In accordance with the Laurent series of reflection coefficient and oscillation term, the determinant formula of the high-order pole solution with NZBCs is established. Finally, combined with specific parameters, the dynamic propagation behaviors of the high-order pole solutions are further analyzed and some very interesting phenomena are obtained, including kink solution, anti kink solution, rational solution and breathing-soliton solution.
\end{abstract}


\end{frontmatter}

\tableofcontents

\section{Introduction}
The nonlinear Schr\"{o}dinger equation
\begin{align}
iq_{t}(x,t)=q_{xx}\pm2|q(x,t)|^2q(x,t)
\end{align}
has been widely concerned and studied in the past until now since the significant work of Zakharov and Shabat \cite{JETP-1972}. As a general model for the evolution of complex envelope of weakly nonlinear dispersive wave train \cite{JMP-1967}, it has very important physical significance in  plasma waves \cite{PRL-2011}, Bose Einstein condensation \cite{NC-1961,JETP-1961} and other physical phenomena. Different from the case where the equation is local, the nonlinear induced ``potential" is parity-time ($\mathcal {P}\mathcal {T}$) symmetry. Since Bender et al. introduced $\mathcal {P}\mathcal {T}$ symmetry into generalized Hamiltonian in 1998 \cite{PRL-1998}, it has played an important role in many fields \cite{RPP-2007,PTRSA-2013,RMP-2016}. In 2013, $\mathcal {P}\mathcal {T}$ symmetry was introduced into the first Ablowitz-Kaup-Newell-Segur (AKNS) type system, and a nonlocal nonlinear Schr\"{o}dinger equation
\begin{align}
iq_{t}(x,t)=q_{xx}\pm2q(x,t)q^*(-x,t)q(x,t)
\end{align}
was proposed by Ablowitz and Musslimani with $q^*(-x,t)$ (a nonlocal term), which  is an integrable system with Lax pair and can be solved by inverse scattering transform (IST) \cite{PRL-2013}. Subsequently, a large number work has been carried out on the study of nonlocal NLS by IST, the properties and interactions of soliton solutions have been studied \cite{Non-2017,Chaos-2017,Non-2017-1,PRE-2015-1}. Gradually, nonlocal equations have attracted many scholars' discussion and research in recent years \cite{Chaos-2019,JMAA-2017,PRE-2015,PRE-2014}. In the nonlinear integrable model, NLS equation and KdV equation are the most basic models, which have high research value. Therefore, we consider the mKdV equation with $\mathcal {P}\mathcal {T}$ symmetry in this work.
For the  mKdV equation
\begin{align}
q_t(x,t)+6q^2(x,t)q_x(x,t)+q_{xxx}(x,t)=0,
\end{align}
which also has very important physical significance \cite{OQE-1998,PS-2006,PRE-1995,Springer-1986,1982,SIAM-1981}. It is particularly worth mentioning that the author used the inverse scattering transform (IST) to study the mKdV equation and obtained the exact soliton solution\cite{JPSJ-1681}. In addition, the solutions derived from the $\mathcal {P}\mathcal {T}$ symmetry potential are also obtained  in \cite{JPSJ-2008}. Subsequently, inspired by Ablowitz and Musslimani's work, Zhu et al. proposed and studied the following nonlocal mKdV equation \cite{CNSNS-2017}
\begin{align}
q_{t}(x,t)+6q(x,t)q(-x,-t)q_x(x,t)+q_{xxx}(x,t)=0,
\end{align}
which can be derived from AKNS system by a reduction.

Some nonlocal integrable nonlinear wave equations with reverse-space-time and reverse-time are proposed, and the IST are further employed to investigate the soliton solution  with zero boundary conditions (ZBCs) of  these  nonlinear wave equations \cite{Non-2016,SIAM-2017}. In 2018, Ablowize et al. introduced a single-valued variable to solve the inverse problem on the standard complex $z$-plane, and presented a framework for the IST of nonlocal equations with nonzero boundary conditions (NZBCs) \cite{JMP-2018,TMP-2018,JMP-2016,IP-2007,SAM-2013,JMP-2014}. Then, the IST method of focusing and defocusing mKdV equation  and the derivative NLS equation with NZBCs were studied at infinity in 2019 \cite{ZGQ-2018,YZY-2018}. Inspired by Yan's work, we employ the RH method to study the integrable real nonlocal mKdV equation \cite{PD-2019}
\begin{align}\label{Eq1}
q_{t}(x,t)-6\sigma q(x,t)q(-x,-t)q_x(x,t)+q_{xxx}(x,t)=0,
\end{align}with NZBCs as follows:
\begin{align}
\lim_{x\rightarrow\pm\infty}q(x,t)=q_{\pm}, ~|q_{\pm}|=q_0>0,
\end{align}
where $\sigma=\pm1$ denote the defocusing and  focusing cases, respectively, and $q_+=\delta q_-$ with $\delta=\pm1$. The potential function $q(x,t)$ is a real function. Therefore, the difference of $\delta$ will lead to the difference of phase. Yan's work has presented a consistent IST to study the soliton solution with NZBCs, including four cases for considering nonlocal integrable nonlinear systems \cite{PD-2019}.

For nonlinear integrable systems, it is generally known  that soliton solutions are generated at the poles of the reflection coefficient $r(z)$, and most integrable systems have high-order poles. For the soliton solutions of nonlocal mKdV equations, Zhu's work uses classical IST and  constructs GLM integral equation  to recover the solution with ZBCs \cite{CNSNS-2017}, while Yan's work uses RH method to study the simple pole and second-order poles with NZBCs \cite{PD-2019}.  It is obviously that  the reflection coefficient has high-order poles naturally can not be ignored. However, it is found that the residue conditions corresponding to the second-order poles needs to be given explicitly when solving the RH problem \cite{JMP-2014,ZGQ-2018,PD-2019,IMAJAM}, and the calculation of  residue conditions with the second-order poles are very complex, even not to  mention the $N$-order poles and multiple high-order poles. Fortunately, the high-order pole solutions with ZBCs \cite{SAM-2020, PD-2019-1} and NZBCs \cite{GI} are obtained by Laurent expansion and Taylor expansion, so as to avoid the complexity of calculating the residue conditions of each order pole. Therefore, we extend the idea in \cite{SAM-2020,PD-2019-1,GI} and apply it  to consider the high-order soliton solution of the nonlocal focusing and defocusing mKdV equation with NZBCs by introducing the generalized residue formula, then some interesting phenomena are obtained, including kink solution, anti kink solution, rational solution and breathing-soliton solution.
%

The frame of the work is arranged as:  In sections 2 and 3, the conditions needed to establish the
corresponding RH problem and the establishment of RH problem are briefly described. The case
of reflection coefficient with a higher-order pole is discussed in section 4, and the corresponding
soliton solution expression is obtained. The case of reflection coefficient with multiple higher-order poles is discussed, and the concrete expression of the solution is given in section 5. Finally, some conclusions and discussions are presented in the last section.
\section{Direct scattering problem}
\subsection{Preliminaries: Lax pair, Riemann surface, and uniformization coordinate}
The nonlocal mKdV equation with $\sigma=\pm1$ are the compatibility condition of the Lax pair
\begin{align}\label{Lax1}
\Phi_{x}=X\Phi,\qquad  \Phi_{t}=T\Phi,
\end{align}
with \begin{subequations}
\begin{align}
&X=X(x,t,k)=ik\sigma_{3}+Q,\\
T=T(x,t,k)=[4k^{2}+2\sigma &q(x,t)q(-x,-t)]X-2ik\sigma_{3}Q_{x}+[Q_{x},Q]-Q_{xx},
\end{align}
\end{subequations}
where $\Phi=\Phi(x,t,k)$ is a matrix eigenfunction, $k$ is a spectral parameter,  the potential matrix $Q$ and the Pauli matrix $\sigma_{3}$ are
\begin{align*}
Q=Q(x,t)=\left(
           \begin{array}{cc}
             0 & q(x,t) \\
             \sigma q(-x,-t) & 0 \\
           \end{array}
         \right),\quad\sigma=\pm1,
\end{align*}
and the Pauli matrices given by
\begin{align*}
\sigma_{1}=\left(
           \begin{array}{cc}
             0 & 1 \\
             1 & 0 \\
           \end{array}
         \right),~~~
\sigma_{2}=\left(
           \begin{array}{cc}
             0 & -i \\
             i & 0 \\
           \end{array}
         \right),~~~
\sigma_{3}=\left(
           \begin{array}{cc}
             1 & 0 \\
             0 & -1 \\
           \end{array}
         \right).
\end{align*}

The asymptotic scattering problem is $\Phi_{x}=X_{\pm}\Phi,$  where $X_{\pm}=ik\sigma_{3}+Q_{\pm}$. The eigenvalues of $X_{\pm}(x,t,k)$ are $\pm i\sqrt{k^{2}-\delta\sigma q_{0}^{2}}.$ Since the eigenvalues are doubly branched, we have introduced the two-sheeted  Riemann surfaces defined by $\lambda^{2}=k^{2}-\delta\sigma q_{0}^{2}$ corresponding to $\delta\sigma=1$ and $\delta\sigma=-1$ so that $\lambda(k)$ is single-valued on this surface. The branch points are the values of $k$ for which $k^{2}-\delta\sigma q_{0}^{2}=0$, i.e., $k=\pm\sqrt{\delta\sigma}q_{0}$. (In the case, $\delta\sigma=1$, and the branch points are $k=\pm q_{0}$; In the case, $\delta\sigma=-1$, and the branch points are $k=\pm iq_{0}$.) Letting $k-q_{0}=r_{1}e^{i\theta_{1}}$ and $k+q_{0}=r_{2}e^{i\theta_{2}}$, then we can write $\lambda_{m}(k)=\sqrt{r_{1}r_{2}}e^{i((\theta_{1}+\theta_{2})/2+m\pi)}(m=0,1)$,  which  is located in sheet-I and sheet-II (In the case, $\delta\sigma=1$, we now take  $\theta_{1,2}\in[-\pi,\pi)$; In the case, $\delta\sigma=-1$, we now take  $\theta_{1,2}\in[-\frac{\pi}{2},\frac{3\pi}{2})$), respectively. With this conventions, the discontinuity of $\lambda(k)$ occurs on the segment are $[-q_{0},q_{0}]$ and $[-iq_{0},iq_{0}]$ corresponding to the two cases $\delta\sigma=1$ and $\delta\sigma=-1$. In order to seek for the analytical regions of the Jost solutions and scattering data, we have to determine the regions where $Im \lambda(k)>0(<0)$. To begin with, from the definition of the two-sheeted Riemann surface, one obtains that the region  where $Im \lambda(k)<0$ is the lower-half plane on Sheet-I and upper-half plane  on Sheet II and $Im \lambda(k)>0$ is the upper-half plane on Sheet-I and the lower-half plane on Sheet-II. The square root sign denotes the principal branch of the real-valued square root function.

In what follows, we take a uniformization variable
\begin{align*}
z=k+\lambda,
\end{align*}
the inverse transformation is
\begin{align*}
k=\frac{1}{2}\left(z+\delta\sigma\frac{q_{0}^{2}}{z}\right),\quad \lambda=\frac{1}{2}\left(z-\delta\sigma\frac{q_{0}^{2}}{z}\right).
\end{align*}

For convenient,  letting
\begin{align}
D^{+}=\left\{ \begin{aligned}
         &\mathbb{C}^{+},
         \quad &\delta\sigma=1,\\
         &\{z\in\mathbb{C}|(|z|-q_{0})Im~z>0\},
         \quad &\delta\sigma=-1,
                          \end{aligned} \right.\\
D^{-}=\left\{ \begin{aligned}
         &\mathbb{C}^{-},
         \quad &\delta\sigma=1,\\
         &\{z\in\mathbb{C}|(|z|-q_{0})Im~z<0\},
         \quad &\delta\sigma=-1,
                          \end{aligned} \right.
\end{align}
The two domains  depicted in Fig. 1. In what follows, these property determines the analyticity regions of the Jost eigenfunctions and means that  the scattering problem on the standard $z$-plane can be discussed instead of on the two-sheet Riemann surface via inverse mapping.

\centerline{\begin{tikzpicture}[scale=0.6]
\path [fill=gray](9,9) -- (9,1) to (1,1) -- (1,9);
\path (-9,1) -- (-9,9) to (-1,9) -- (-1,1);
\path [fill=gray] (-9,5)--(-9,9) to (-1,9) -- (-1,5);
\filldraw (-9,1) -- (-9,9) to (-1,9) -- (-1,1);
\filldraw (9,9) -- (9,1) to (1,1) -- (1,9);
\filldraw[white, line width=0.5](-1,5)--(3,5) arc (-180:0:2);
\path [fill=gray] (1,5) -- (9,5) to (9,9) -- (1,9);
\filldraw[white, line width=0.5](-1,5)--(3,5) arc (-180:0:2);
\path [fill=white] (1,1) -- (9,1) to (9,5) -- (1,5);
\path [fill=gray] (-9,5)--(-9,9) to (-1,9) -- (-1,5);
\path [fill=white] (-9,1)--(-9,5) to (-1,5) -- (-1,1);
\filldraw[gray, line width=0.5](1,5)--(3,5) arc (-180:0:2);
\filldraw[white, line width=0.5](1,5)--(3,5) arc (180:0:2);
\draw[fill] (-5,5)node[below]{} circle [radius=0.035];
\draw[fill] (5,5)node[below]{} circle [radius=0.035];
\draw[fill] (-7.5,7.5)node[left]{$-z_{n}^{*}$};
\draw[blue, fill] (-7.5,7.5) circle [radius=0.035]node[below]{};
\draw[fill] (-2.6,7.5)node[right]{$z_{n}$};
\draw[blue, fill] (-2.6,7.5) circle [radius=0.045]node[below]{};
\draw[fill] (-5,7.5)node[left]{$i\omega_{n}$};
\draw[blue, fill] (-5,7.5) circle [radius=0.045]node[below]{};
\draw[fill] (2.5,7.5)node[left]{$-z_{n}^{*}$};
\draw[blue, fill] (2.5,7.5) circle [radius=0.035]node[below]{};
\draw[fill] (7.5,7.5)node[right]{$z_{n}$};
\draw[blue, fill] (5,7.5) circle [radius=0.045]node[below]{};
\draw[fill] (5,7.5)node[left]{$i\omega_{n}$};
\draw[blue, fill] (7.5,7.5) circle [radius=0.035]node[below]{};
\draw[yellow, fill] (-5.5,4.5) circle [radius=0.035]node[below]{};
\draw[fill] (-5.5,4.45)node[left]{$-\frac{q_{0}^{2}}{z_{n}^{*}}$};
\draw[yellow, fill] (-4.5,4.5) circle [radius=0.035]node[below]{};
\draw[fill] (-4.5,4.45)node[right]{$\frac{q_{0}^{2}}{z_{n}}$};
\draw[yellow, fill] (-5,4.5) circle [radius=0.1]node[below]{};
\draw[fill] (-5,3.7)node[left]{$-\frac{iq_{0}^{2}}{\omega_{n}}$};
\draw[yellow, fill] (5.5,5.5) circle [radius=0.035]node[below]{};
\draw[fill] (5.5,5.5)node[right]{$\frac{q_{0}^{2}}{z_{n}^{*}}$};
\draw[yellow, fill] (4.5,5.5) circle [radius=0.035]node[below]{};
\draw[fill] (4.5,5.5)node[left]{$-\frac{q_{0}^{2}}{z_{n}}$};
\draw[yellow, fill] (5,5.5) circle [radius=0.1]node[below]{};
\draw[fill] (5,6.5)node[right]{$\frac{iq_{0}^{2}}{\omega_{n}}$};
\draw[fill] (-7,5) circle [radius=0.055]node[below]{\footnotesize$-q_{0}$};
\draw[fill] (-3,5) circle [radius=0.055]node[below]{\footnotesize$q_{0}$};
\draw[-][black, line width=0.5][thick](1,5)--(9,5);
\draw[-][thick](-9,5)--(-8,5)[->];
\draw[-][thick](-7,5)--(-6,5)[->];
\draw[-][thick](-5,5)--(-4,5)[->];
\draw[-][thick](-3,5)--(-2,5)[->];
\draw[-][thick](1,5)--(2,5)[->];
\draw[-][thick](3,5)--(4,5)[-<];
\draw[-][thick](5,5)--(6,5)[-<];
\draw[-][thick](7,5)--(8,5)[->];
\draw[-][thick](-9,5)--(-8,5);
\draw[-][thick](-8,5)--(-7,5);
\draw[-][thick](-7,5)--(-6,5);
\draw[-][thick](-6,5)--(-5,5);
\draw[-][thick](-5,5)--(-4,5);
\draw[-][thick](-4,5)--(-3,5);
\draw[-][thick](-3,5)--(-2,5);
\draw[-][thick](-2,5)--(-1,5)[->][thick]node[above]{$Rez$};;
\draw[-][thick](-5,1)--(-5,2);
\draw[-][thick](-5,2)--(-5,3);
\draw[-][thick](-5,3)--(-5,4);
\draw[-][thick](-5,4)--(-5,5);
\draw[-][thick](-5,5)--(-5,6);
\draw[-][thick](-5,6)--(-5,7);
\draw[-][thick](-5,7)--(-5,8);
\draw[-][thick](-5,8)--(-5,9)[->] [thick]node[above]{$Imz$};
\draw[-][thick](1,5)--(2,5);
\draw[-][thick](2,5)--(3,5);
\draw[-][thick](3,5)--(4,5);
\draw[-][thick](4,5)--(5,5);
\draw[-][thick](5,5)--(6,5);
\draw[-][thick](6,5)--(7,5);
\draw[-][thick](7,5)--(8,5);
\draw[-][thick](8,5)--(9,5)[->][thick]node[above]{$Rez$};
\draw[-][thick](5,1)--(5,2);
\draw[-][thick](5,2)--(5,3);
\draw[-][thick](5,3)--(5,4);
\draw[-][thick](5,4)--(5,5);
\draw[-][thick](5,5)--(5,6);
\draw[-][thick](5,6)--(5,7);
\draw[-][thick](5,7)--(5,8);
\draw[-][thick](5,8)--(5,9);
\draw[->](5,9)[thick]node[above]{$Imz$};
\draw[fill] (-5,7) circle [radius=0.055]node[left]{};
\draw[fill] (-5,3) circle [radius=0.055]node[left]{};
\draw[fill] (5,7) circle [radius=0.055]node[above]{};
\draw[fill] (5,3) circle [radius=0.055]node[below]{};
\draw[fill](5.5,7)[thick]node[above]{\footnotesize$iq_{0}$};
\draw[fill](5.5,3)[thick]node[below]{\footnotesize$-iq_{0}$};
\draw[black, line width=0.5](-9,5)--(-7,5) arc (-180:0:2);
\draw[black, line width=0.5](-9,5)--(-7,5) arc (180:0:2);
\end{tikzpicture}}
\noindent { \small \textbf{Figure 1.}  The grey $(D_{+})$ and white $(D_{-})$ regions stand for $Im\lambda(z)>0$ and $Im\lambda(z)<0$, respectively. The  complex $z$-plane for $\delta\sigma=1$(left) and $\delta\sigma=-1$ (right) showing the discrete spectrums [zeros of
scattering data $s_{11}(z)$ (blue) in grey region and those of scattering data $s_{22}(z)$ (yellow) in white region], and the orientation of the contours for the Riemann-Hilbert problem.}

\subsection{Jost solutions and analyticity}
The asymptotic scattering problem $(x\rightarrow\pm\infty)$ of the original Lax pair \eqref{Lax1} are considered as
\begin{align}
\Phi_{x}=X_{\pm}(k)\Phi,\quad \Phi_{t}=T_{\pm}(k)\Phi,
\end{align}
where $X_{\pm}=ik\sigma_{3}+Q_{\pm}$ and $T_{\pm}=(4k^{2}+2\sigma\delta q_{0}^{2})X_{\pm}$ with $Q_{\pm}
=\left(
    \begin{array}{cc}
      0 & q_{\pm} \\
      \sigma\delta q_{\pm} & 0 \\
        \end{array}
         \right),$
the fundamental matrix solution are obtained , that is
\begin{align}
\Phi_{\pm}(x,t,k)=\left\{ \begin{aligned}
 &E_{\pm}(k)e^{i\theta(x,t,k)\sigma_{3}},~\qquad\qquad\qquad\quad k\neq\sqrt{\delta\sigma} q_{0},\\
         &I+[x+(4k^{2}+2\sigma\delta q_{0}^{2})t]X_{\pm}(k),
         \quad k=\sqrt{\delta\sigma} q_{0},
\end{aligned} \right.
\end{align}
where
\begin{align}
\begin{split}
E_{\pm}(k)&=\left(
             \begin{array}{cc}
               1 & \frac{iq_{\pm}}{k+\lambda} \\
               -\delta\sigma\frac{iq_{\pm}}{k+\lambda} & 1 \\
             \end{array}
           \right),\\
          \theta(x,t,k)=\lambda[x+&(4k^{2}+2\delta\sigma q_{0}^{2})t], \quad \lambda^{2}=k^{2}-\delta\sigma q_{0}^{2}.
          \end{split}
\end{align}

We define the continuous spectrum as $\Sigma$, correspondingly,  $\lambda(z)\in\mathbb{R}$ exactly for $z\in\Sigma$. Then we define
\begin{align}
\Sigma=\left\{ \begin{aligned}
         &\mathbb{R},
         \qquad\qquad\qquad\qquad \delta\sigma=1,\\
         &\mathbb{R}\cup\{z\in\mathbb{C}:|z|=q_{0}\},
         \quad \delta\sigma=-1.
                          \end{aligned} \right.
\end{align}
For all $z\in\Sigma$, we  define the Jost eigenfunctions $\Phi_{\pm}(x,t,k)$ as the simultaneous solutions of both parts of the Lax pair such that
\begin{align}
\Phi_{\pm}(x,t,z)=E_{\pm}(z)e^{i\theta(x,t,z)\sigma_{3}}+O(1),\quad  x\rightarrow\pm\infty.
\end{align}
We can subtract the asymptotic state of the potential and rewrite the first one in \eqref{Lax1} as
\begin{align}
\Phi_{\pm,x}=X_{\pm}(z)\Phi_{\pm}+\Delta Q_{\pm}\Phi_{\pm},
\end{align}
where $Q_{\pm}(x,t)=Q-Q_{\pm}$. For simplicity, the modified Jost solutions $\mu_{\pm}(x,t,z)$ are introduced in order to eliminating the exponential oscillations:
\begin{align}\label{Jost1}
\mu_{\pm}(x,t,z)=\Phi_{\pm}(x,t,z)e^{-i\theta(x,t,z)\sigma_{3}},
\end{align}
so that $\mathop{\lim}\limits_{x\rightarrow\pm\infty}\mu_{\pm}(x,t,z)=E_{\pm}(x,t,z)$.
The ordinary differential equation for $\mu_{\pm}(x,t,z)$ can be integrated to obtain linear integral equations for the modified Jost solutions:
\begin{align}\label{In1}
\mu_{\pm}(x,t,z)=E_{\pm}(z)+\left\{ \begin{aligned}
         &\int_{\pm\infty}^{x}E_{\pm}(z)e^{i\lambda(z)(x-y)\hat{\sigma}_{3}}[E_{\pm}^{-1}(z)
         \Delta Q_{\pm}(y,t)\mu_{\pm}(y,t,z)]dy,
         ~~ z\neq{\sqrt\delta\sigma}q_{0},\\
         &\int_{\pm\infty}^{x}[I+(x-y)X_{\pm}(z)]\Delta Q_{\pm}(y,t)\mu_{\pm}(y,t,z)dy,
         ~~~~~~~~~\quad z={\sqrt\delta\sigma}q_{0},
                          \end{aligned} \right.
\end{align}
where $e^{i\lambda(z)(x-y)\hat{\sigma}_{3}}A=e^{i\lambda(z)(x-y)\hat{\sigma}_{3}}A
e^{-i\lambda(z)(x-y)\hat{\sigma}_{3}}$. In what follows, for convenience, we let $\Sigma_{0}=\Sigma\setminus\{\pm iq_{0}\}$ as $\delta\sigma=-1$ and $\Sigma_{0}=\Sigma\setminus\{\pm q_{0}\}$ as $\delta\sigma=1$.
\begin{prop}
Providing that $q(x,t)-q_{\pm}\in L^{1}(\mathbb{R}^{\pm})$, the Votelrra integral equation \eqref{In1} satisfies unique solutions $\mu_{\pm}(x,t,z)$ defined by \eqref{Jost1} in $\Sigma_{0}$. In addition, the columns $\mu_{-,1}(x,t,z)$ and $\mu_{+,2}(x,t,z)$ can be analytically extended to $D^{+}$ and continuously extended to $D^{+}\cup\Sigma_{0}$, while the columns $\mu_{+,1}(x,t,z)$ and $\mu_{-,2}(x,t,z)$ can be analytically extended to $D^{-}$ and continuously extended to $D^{-}\cup\Sigma_{0}$, where $\mu_{\pm,j}(x,t,z)$ denote the $j$-th column of $\mu_{\pm}(x,t,z)$.
\end{prop}

\begin{prop}
Providing that $(1+|x|)(q(x,t)-q_{\pm})\in L^{1}(\mathbb{R}^{\pm})$, the Votelrra integral equation \eqref{In1} have unique solutions $\mu_{\pm}(x,t,z)$ defined by \eqref{Jost1} in $\Sigma$. In addition, the columns $\mu_{-,1}(x,t,z)$ and $\mu_{+,2}(x,t,z)$ can be analytically extended to $D^{+}$ and continuously extended to $D^{+}\cup\Sigma$, while the columns $\mu_{+,1}(x,t,z)$ and $\mu_{-,2}(x,t,z)$ can be analytically extended to $D^{-}$ and continuously extended to $D^{-}\cup\Sigma$.
\end{prop}

\begin{lem}
The Jost solutions $\Phi(x,t,z)$  is also the simultaneous solution of Lax pair \eqref{Lax1}, then $\Phi(x,t,z)$  and $\mu(x,t,z)$  have the same analytic properties.
\end{lem}

\subsection{Scattering matrix}
It follows from the Liouville's formula that $\Phi_{\pm}(x,t,z)$ are fundamental solutions of the Lax pair \eqref{Lax1} as $z\in\Sigma_{0}$, then there exists a scattering matrix $S(z)$, such that
\begin{align}\label{sr}
\Phi_{+}(x,t,z)=\Phi_{-}(x,t,z)S(z),\quad x,t\in\mathbb{R},~~z\in\Sigma_{0}.
\end{align}
(It is obviously that one could equivalently write $\Phi_{-}(x,t,z)=\Phi_{+}(x,t,z)R(z)$, which also is the traditional way). Columuwise, and we have the Wranskian determinant of the scattering coefficients $s_{ij}(z)(i,j=1,2)$
\begin{align*}
s_{11}(z)=\frac{Wr(\Phi_{+,1}(x,t,z),\Phi_{-,2}(x,t,z))}{1-\delta\sigma q_{0}^{2}/z^{2}},~~
s_{12}(z)=\frac{Wr(\Phi_{+,2}(x,t,z),\Phi_{-,2}(x,t,z))}{1-\delta\sigma q_{0}^{2}/z^{2}},\\
s_{21}(z)=\frac{Wr(\Phi_{-,1}(x,t,z),\Phi_{+,1}(x,t,z))}{1-\delta\sigma q_{0}^{2}/z^{2}},~~
s_{22}(z)=\frac{Wr(\Phi_{-,1}(x,t,z),\Phi_{+,2}(x,t,z))}{1-\delta\sigma q_{0}^{2}/z^{2}}.
\end{align*}

\begin{prop}
Providing that $q(x,t)-q_{\pm}\in L^{1}(\mathbb{R}^{\pm})$, the scattering coefficients $s_{11}(z)$ is analytic in $D^{+}$ and continuously in $D^{+}\cup\Sigma_{0}$ while $s_{22}(z)$ is analytic in $D^{-}$ and continuously in $D^{-}\cup\Sigma_{0}$. As usual, the off-diagonal scattering coefficients are nowhere analytic in general, both $s_{12}(z)$ and $s_{21}(z)$ are continuous in $\Sigma_{0}$.
\end{prop}

\begin{prop}
Providing that $(1+|x|)(q(x,t)-q_{\pm})\in L^{1}(\mathbb{R}^{\pm})$, the scattering coefficients $s_{11}(z)$ is analytic in $D^{+}$ and continuously in $D^{+}\cup\Sigma$ while $s_{22}(z)$ is analytic in $D^{-}$ and continuously in $D^{-}\cup\Sigma$. Moreover, both $\lambda(z)s_{12}(z)$ and $\lambda(z)s_{21}(z)$ are continuous in $\Sigma$.
\end{prop}
The reflection coefficients that will be needed in the inverse problem are
\begin{align}
r(z)=\frac{s_{21}(z)}{s_{11}(z)},\quad \widetilde{r(z)}=\frac{s_{12}(z)}{s_{22}(z)},\quad z\in\Sigma.
\end{align}

\subsection{Symmetries}
The symmetry properties of the Jost solutions and scattering matrix are studied. In what follows, the three symmetry relations are derived. For convenient, we define $\sigma_{*}=\left\{ \begin{aligned}
         &\sigma_{1},
         \quad \sigma=-1,\\
         &\sigma_{2},
         \quad \sigma=1,
                          \end{aligned} \right.$ The matrix $\sigma_{*}$ is defined now
for later use.\\

\begin{enumerate}[(i)]
\item The first symmetry: $X(x,t,z)$ and $T(x,t,z)$ in \eqref{Lax1},  the Jost solutions $\Phi_{\pm}(x,t,z)$ and the scattering matrix $S(z)$ satify the relations:
\begin{subequations}
\begin{align}
X(x,t,z)&=-\sigma_{*}X(-x,-t,-z^{*})\sigma_{*},\quad T(x,t,z)=-\sigma_{*}T(-x,-t,-z^{*})\sigma_{*},\\
\Phi_{\pm}(x,t,z)&=-\sigma_{*}\Phi_{\mp}^{*}(-x,-t,-z^{*})\sigma_{*},~~~S(z)=-\sigma_{*}[S^{*}(-z^{*})
]^{-1}\sigma_{*}.
\end{align}
\end{subequations}

\item The second symmetry: $X(x,t,z)$ and $T(x,t,z)$ in \eqref{Lax1},  the Jost solutions $\Phi_{\pm}(x,t,z)$ and the scattering matrix $S(z)$ satisfy the relations:
\begin{subequations}
\begin{align}
X(x,t,z)&=X^{*}(x,t,-z^{*}),\quad T(x,t,z)=T^{*}(x,t,-z^{*}),\\
\Phi_{\pm}(x,t,z)&=\Phi_{\pm}^{*}(x,t,-z^{*}),\quad S(z)=[S^{*}(-z^{*})].
\end{align}
\end{subequations}

\item The third symmetry: $X(x,t,z)$ and $T(x,t,z)$ in \eqref{Lax1},  the Jost solutions $\Phi_{\pm}(x,t,z)$ and the scattering matrix $S(z)$ satisfy the relations:
\begin{subequations}
\begin{align}
X(x,t,z)&=X^{*}\left(x,t,\delta\sigma\frac{q_{0}^{2}}{z}\right),\quad ~ T(x,t,z)=T^{*}\left(x,t,\delta\sigma\frac{q_{0}^{2}}{z}\right),\\
\Phi_{\pm}(x,t,z)&=\frac{i}{z}\Phi_{\pm}^{*}\left(x,t,\delta\sigma\frac{q_{0}^{2}}{z}\right),\quad S(z)=(\sigma_{3}Q_{-})^{-1}S\left(\delta\sigma\frac{q_{0}^{2}}{z}\right)(\sigma_{3}Q_{+}).
\end{align}
\end{subequations}
\end{enumerate}

\subsection{Asymptotic behaviors}
Normally the asymptotic properties of the modified Jost solutions and scattering matrix are needed to define the inverse problem properly. Similarly, the asymptotic behavior of NZBCs is more complex than that of ZBCs, but it is simplified in the uniformization variable.
\begin{lem}
The asymptotic behaviors for the modified Jost solutions and the scattering matrix are derived by
\begin{align}
\mu_{\pm}(x,t,z)=\left\{ \begin{aligned}
         &I+O\left(z^{-1}\right),
         \quad z\rightarrow\infty,\\
         &\frac{i}{z}\sigma_{3}Q_{\pm}+O(1),
         \quad z\rightarrow0,
                          \end{aligned} \right.
\end{align}
and
\begin{align}
S(z)=\left\{ \begin{aligned}
         &I+O\left(z^{-1}\right),
         \quad z\rightarrow\infty,\\
         &\delta I+O(z),
         \quad z\rightarrow0.
                          \end{aligned} \right.
\end{align}
\end{lem}
We can easily derive that
\begin{align}\label{q1}
q(x,t)=-\lim_{z\rightarrow\infty}i(z\mu_{\pm})_{12}.
\end{align}
\subsection{Discrete spectrum}
As usual, the discrete spectrum of the scattering problem is the set of all values $z\in\mathbb{C}$
such that eigenfunctions exist in $L^{2}\mathbb(R)$. We next show that these values are the zeros of $s_{11}(z)$ in $D^{+}$ and those of $s_{22}(z)$ in $D^{-}$. We suppose that $s_{11}$ has $N$-th order zero $z_{0}$ in $D^{+}\cap\{z\in\mathbb{C}: Re~z>0\}$. It follows from the symmetry relations of the scattering coefficients that
\begin{align}
s_{11}(z_{0})=s_{11}(-z_{0}^{*})=s_{22}\left(\delta\sigma\frac{q_{0}^{2}}{z_{0}}\right)=
s_{22}\left(-\delta\sigma\frac{q_{0}^{2}}{z_{0}^{*}}\right).
\end{align}
To this end, the discrete spectrum is given by
\begin{align}
Z=\left\{z_{0}, -z_{0}^{*}, \delta\sigma\frac{q_{0}^{2}}{z_{0}}, -\delta\sigma\frac{q_{0}^{2}}{z_{0}^{*}}\right\},
\end{align}
which can be seen in Figure 1.

\section{Riemann-Hilbert problem}
With normal circumstances, we start from the scattering relationship to find the formula of the inverse problem, that is, the relationship between the eigenfunctions analytic in $D^{+}$ and the eigenfunctions analytic in $D^{-}$. We introduce  piecewise meromorphic matrix as follows:
\begin{align}\label{rh}
M^{+}(x,t,z)=\left(\frac{\mu_{+,1}}{s_{11}}, \mu_{-,2}\right),\quad M^{-}(x,t,z)=\left(\mu_{-,1}, \frac{\mu_{+,2}}{s_{22}}\right),
\end{align}
the jump condition is
\begin{align*}
M^{-}(x,t,z)=M^{+}(x,t,z)(I-G(x,t,z)), \quad z\in\Sigma,
\end{align*}
where the jump matrix
\begin{align}
G(x,t,z)=\left(
           \begin{array}{cc}
             0 & -e^{2i\theta}\widetilde{\rho} \\
             e^{-2i\theta}\rho & \rho\widetilde{\rho} \\
           \end{array}
         \right).
\end{align}
Recall that the asymptotic behavior of the Jost eigenfunctions and the scattering coefficients, it is straightforward to  check that
\begin{align}
M(x,t,z)=\left\{ \begin{aligned}
         &I+O\left(z^{-1}\right),
         \quad z\rightarrow\infty,\\
         &\frac{i}{z}\sigma_{3}Q_{-}+O(1),
         \quad z\rightarrow0.
                          \end{aligned} \right.
\end{align}
From \eqref{q1}, we can get that
\begin{align}
q(x,t)=-\lim_{z\rightarrow\infty}i\left(zM\right)_{12}.
\end{align}

\section{Single high-order pole solutions}
Let $z_{0}\in D^{+}$ is the $N$th-order pole, because of the symmetry properties it is  straightforward that $-z_{0}^{*}, \frac{q_{0}^{2}}{z_{0}}, -\frac{q_{0}^{2}}{z_{0}^{*}}\in D^{+}$ also is the $N$th-order pole of $s_{11}(z)$. The discrete spectrum is the set
\begin{align*}
\left\{z_{0}, -z_{0}^{*},  \frac{q_{0}^{2}}{z_{0}}, -\frac{q_{0}^{2}}{z_{0}^{*}}\right\},
\end{align*}
which shown in Figure 1.

Let $v_{1}=z_{0}, v_{1}=\frac{q_{0}^{2}}{z_{0}}$, and
\begin{align}
s_{11}(z)=(z^{2}-v_{1}^{2})^{N}(z^{2}-v_{2}^{2})^{N}s_{0}(z),
\end{align}
in which $s_{0}(z)\neq0\in D^{+}$. Developed at the pole points based on the Laurent expansion, $r(z)$ and $\widetilde{r}(z)$ can be expand respectively as
\begin{align}\label{rz}
r(z)=r_{0}^{1}(z)+\sum_{n=1}^{N}\frac{r_{1,n}^{1}}{(z-v_{1})^{n}},\quad\quad
r(z)=r_{0}^{2}(z)+\sum_{n=1}^{N}\frac{r_{1,n}^{2}}{(z+v_{1}^{*})^{n}}(-1)^{n+1},\\
\widetilde{r(z)}=\widetilde{r_{0}^{1}(z)}+\sum_{n=1}^{N}\frac{r_{2,n}^{1}}{(z-v_{2})^{n}},\quad\quad
\widetilde{r(z)}=\widetilde{r_{0}^{2}(z)}+\sum_{n=1}^{N}\frac{r_{2,n}^{2}}{(z+v_{2}^{*})^{n}}(-1)^{n+1},
\end{align}
where $r_{m}$ can be written as
\begin{align*}
r_{j,n}^{\ell}=\lim_{z\rightarrow v_{j}}\frac{1}{(N-n)!}\frac{\partial^{N-n}}{\partial z^{N-n}}
\left[(z-v_{j})^{N}r(z)\right],  n=1,...,N,
\end{align*}
and $r_{0}^{1}(z)$ and $r_{0}^{2}(z)$ are analytic for all $z\in D^{+}$. However, because of the definition of the matrix function $M(x,t,z)$, we know that $M_{11}$ has $N$th-order poles at $z=v_{1}$ and $z=-v_{1}^{*}$, while  $M_{12}$ has $N$th-order poles at $z=v_{2}$ and $z=-v_{2}^{*}$. From the normalization condition, one has
\begin{align}\label{M12}
\begin{split}
M_{11}(x,t,z)&=1+\sum_{s=1}^{N}\left(\frac{H_{s}(x,t)}{(z-v_{1})^{s}}+
\frac{J_{s}(x,t)}{(z+v_{1}^{*})^{s}}\right),\\
M_{12}(x,t,z)&=\frac{iq_{-}}{z}+\sum_{s=1}^{N}\left(\frac{K_{s}(x,t)}{(z-v_{2})^{s}}+
\frac{L_{s}(x,t)}{(z+v_{2}^{*})^{s}}\right),
\end{split}
\end{align}
where $H_{s}(x,t), J_{s}(x,t), K_{s}(x,t)$ and $L_{s}(x,t)(s=1,\ldots, N)$ are  to-be-determined functions. In order to obtain the expressions of $H_{s}(x,t), J_{s}(x,t), K_{s}(x,t),$ and $L_{s}(x,t)(s=1,\ldots, N)$, we take the Taylor series expansion
\begin{subequations}
\begin{align}
e^{-2i\theta}&=\sum_{\ell=0}^{\infty}f_{1,\ell}^{1}(z-v_{1})^{\ell},\quad
e^{-2i\theta}=\sum_{\ell=0}^{\infty}f_{1,\ell}^{2}(z+v_{1}^{*})^{\ell}(-1)^{\ell},\\
e^{2i\theta}&=\sum_{\ell=0}^{\infty}f_{2,\ell}^{1}(z-v_{2})^{\ell},\quad~~
e^{2i\theta}=\sum_{\ell=0}^{\infty}f_{2,\ell}^{2}(z+v_{2}^{*})^{\ell}(-1)^{\ell},
\end{align}
\end{subequations}
where the elements $f_{j,\ell}^{s}(j=1,2, s=1,2)$ are
\begin{subequations}
\begin{align}
f_{1,\ell}^{1}&=\lim_{z\rightarrow v_{1}}\frac{\partial^{\ell}}{\partial z^{\ell}}e^{-2i\theta},\quad
f_{1,\ell}^{2}=\lim_{z\rightarrow -v_{1}^{*}}\frac{\partial^{\ell}}{\partial z^{\ell}}e^{-2i\theta},\\
f_{2,\ell}^{1}&=\lim_{z\rightarrow v_{2}}\frac{\partial^{\ell}}{\partial z^{\ell}}e^{2i\theta},\quad~~
f_{2,\ell}^{2}=\lim_{z\rightarrow -v_{2}^{*}}\frac{\partial^{\ell}}{\partial z^{\ell}}e^{2i\theta}.
\end{align}
\end{subequations}
The elements $M_{11}$ and $M_{12}$ of the matrix function $M(x,t,z)$ can be written the form
\begin{subequations}
\begin{align}
M_{12}&=\sum_{\ell=0}^{\infty}\xi_{1,\ell}^{1}(z-v_{1})^{\ell},\quad
M_{12}=\sum_{\ell=0}^{\infty}\xi_{1,\ell}^{2}(z+v_{1}^{*})^{\ell}(-1)^{\ell+1},\\
M_{11}&=\sum_{\ell=0}^{\infty}\mu_{2,\ell}^{1}(z-v_{2})^{\ell},\quad~~
M_{11}=\sum_{\ell=0}^{\infty}\mu_{2,\ell}^{2}(z+v_{2}^{*})^{\ell}(-1)^{\ell+1},
\end{align}
\end{subequations}
where
\begin{subequations}\label{f1}
\begin{align}
\xi_{1,\ell}^{1}&=\lim_{z\rightarrow v_{1}}\frac{1}{\ell !}\frac{\partial^{\ell}}{\partial z^{\ell}}M_{12},\quad
\xi_{1,\ell}^{2}=\lim_{z\rightarrow -v_{1}^{*}}\frac{1}{\ell !}\frac{\partial^{\ell}}{\partial z^{\ell}}M_{12},\\
\mu_{2,\ell}^{1}&=\lim_{z\rightarrow v_{2}}\frac{1}{\ell !}\frac{\partial^{\ell}}{\partial z^{\ell}}M_{11},\quad
\mu_{2,\ell}^{2}=\lim_{z\rightarrow -v_{2}^{*}}\frac{1}{\ell !}\frac{\partial^{\ell}}{\partial z^{\ell}}M_{11}.
\end{align}
\end{subequations}
Based on the scattering relation given by \eqref{sr} and the definition of RH given by \eqref{rh}, $H_{s}(x,t)$ and $J_{s}(x,t)$ can be expressed by $\xi_{1,\ell}^{1}$ and $\xi_{1,\ell}^{2}$ via comparing the coefficients of $(z-v_{1})^{-s}$ and $(z+v_{1}^{*})^{-s}$, one has
\begin{subequations}\label{FH}
\begin{align}
H_{s}(x,t)&=\sum_{j=s}^{N}\sum_{\ell=0}^{j-s}r_{1,j}^{1}f_{1,j-s-\ell}^{1}(x,t)\xi_{1,\ell}^{1}(x,t),\\
J_{s}(x,t)&=\sum_{j=s}^{N}\sum_{\ell=0}^{j-s}(-1)^{s}r_{1,j}^{2}f_{1,j-s-\ell}^{2}(x,t)\xi_{1,\ell}^{2}
(x,t).
\end{align}
\end{subequations}
Similarly, by the same method, we can derive that
\begin{subequations}\label{GL}
\begin{align}
K_{s}(x,t)&=\sum_{j=s}^{N}\sum_{\ell=0}^{j-s}r_{2,j}^{1}f_{2,j-s-\ell}^{1}(x,t)\mu_{2,\ell}^{1}(x,t),\\
L_{s}(x,t)&=\sum_{j=s}^{N}\sum_{\ell=0}^{j-s}(-1)^{s}r_{2,j}^{2}f_{2,j-s-\ell}^{2}(x,t)\mu_{2,\ell}^{2}
(x,t).
\end{align}
\end{subequations}
$\xi_{1,\ell}^{t}$ and $\mu_{2,\ell}^{t}$ $(t=1,2)$ actually can also be expressed by $F_{s}, H_{s}, G_{s}$ and $L_{s}$. Reviewing the definition of $\xi_{1,\ell}^{t}$ and $\mu_{2,\ell}^{t}(t=1,2)$ given by \eqref{f1} and substituting \eqref{M12} into them, one has
\begin{subequations}\label{FH1}
\begin{align}
\xi_{1,\ell}^{1}=\frac{(-1)^{\ell}iq_{-}}{(v_{1})^{\ell+1}}+\sum_{s=1}^{N}\left(
                                                                                 \begin{array}{c}
                                                                                   s+\ell+1 \\
                                                                                   \ell \\
                                                                                 \end{array}
                                                                               \right)
\left[\frac{(-1)^{\ell}K_{s}(x,t)}{(v_{1}-v_{2})^{s+\ell}}+
\frac{(-1)^{\ell}L_{s}(x,t)}{(v_{1}+v_{2}^{*})^{s+\ell}}\right],\\
\xi_{1,\ell}^{2}=\frac{(-1)^{\ell}iq_{-}}{(-v_{1}^{*})^{\ell+1}}+\sum_{s=1}^{N}\left(
                                                                                 \begin{array}{c}
                                                                                   s+\ell+1 \\
                                                                                   \ell \\
                                                                                 \end{array}
                                                                               \right)
\left[\frac{(-1)^{\ell}K_{s}(x,t)}{(-v_{1}^{*}-v_{2})^{s+\ell}}+
\frac{(-1)^{\ell}L_{s}(x,t)}{(-v_{1}^{*}+v_{2}^{*})^{s+\ell}}\right],\\
\mu_{2,\ell}^{1}=\left\{ \begin{aligned}
         &1+\sum_{s=1}^{N}\left[\frac{H_{s}(x,t)}{(v_{2}-v_{1})^{s+\ell}}+
\frac{J_{s}(x,t)}{(v_{2}+v_{1}^{*})^{s+\ell}}\right], \quad \ell=0,\\
         &\sum_{s=1}^{N}\left(
                                                                                 \begin{array}{c}
                                                                                   s+\ell+1 \\
                                                                                   \ell \\
                                                                                 \end{array}
                                                                               \right)
\left[\frac{(-1)^{\ell}H_{s}(x,t)}{(v_{2}-v_{1})^{s+\ell}}+
\frac{(-1)^{\ell}J_{s}(x,t)}{(v_{2}+v_{1}^{*})^{s+\ell}}\right],
         \quad \ell=1,2,\cdots,
                          \end{aligned} \right.\\
\mu_{2,\ell}^{2}=\left\{ \begin{aligned}
         &1+\sum_{s=1}^{N}\left[\frac{F_{s}(x,t)}{(-v_{2}^{*}-v_{1})^{s+\ell}}+
\frac{J_{s}(x,t)}{(-v_{2}^{*}+v_{1}^{*})^{s+\ell}}\right], \quad \ell=0,\\
         &\sum_{s=1}^{N}\left(
                                                                                 \begin{array}{c}
                                                                                   s+\ell+1 \\
                                                                                   \ell \\
                                                                                 \end{array}
                                                                               \right)
\left[\frac{(-1)^{\ell}H_{s}(x,t)}{(-v_{2}^{*}-v_{1})^{s+\ell}}+
\frac{(-1)^{\ell}J_{s}(x,t)}{(-v_{2}^{*}+v_{1}^{*})^{s+\ell}}\right],
         \quad \ell=1,2,\cdots,
                          \end{aligned} \right.
\end{align}
\end{subequations}
Using equations \eqref{FH}-\eqref{FH1}, the system can be derived as
\begin{align}\label{FHGL}
\begin{split}
H_{s}(x,t)&=\sum_{j=s}^{N}\sum_{\ell=0}^{j-s}\frac{(-1)^{\ell}iq_{-}}{(v_{1})^{\ell+1}}
r_{1,j}^{1}f_{1,j-s-\ell}^{1}\\&+\sum_{j=s}^{N}\sum_{\ell=0}^{j-s}\sum_{p=1}^{N}\left(
                                                                               \begin{array}{c}
                                                                                 p+\ell+1 \\
                                                                                 \ell \\
                                                                               \end{array}
                                                                             \right)
r_{1,j}^{1}f_{1,j-s-\ell}^{1}\left[\frac{(-1)^{\ell}K_{p}(x,t)}{(v_{1}-v_{2})^{p+\ell}}+
\frac{(-1)^{\ell}L_{p}(x,t)}{(v_{1}+v_{2}^{*})^{p+\ell}}\right],
\end{split}\end{align}
\begin{align}
\begin{split}
J_{s}(x,t)&=\sum_{j=s}^{N}\sum_{\ell=0}^{j-s}\frac{(-1)^{\ell+s}iq_{-}}{(-v_{1}^{*})^{\ell+1}}
r_{1,j}^{1}f_{1,j-s-\ell}^{1}\\&+\sum_{j=s}^{N}\sum_{\ell=0}^{j-s}\sum_{p=1}^{N}\left(
                                                                               \begin{array}{c}
                                                                                 p+\ell+1 \\
                                                                                 \ell \\
                                                                               \end{array}
                                                                             \right)
(-1)^{s}r_{1,j}^{1}f_{1,j-s-\ell}^{1}\left[\frac{(-1)^{\ell}K_{p}(x,t)}{(-v_{1}^{*}-v_{2})^{p+\ell}}+
\frac{(-1)^{\ell}L_{p}(x,t)}{(-v_{1}^{*}+v_{2}^{*})^{p+\ell}}\right],\\
K_{s}(x,t)&=\sum_{j=s}^{N}\sum_{\ell=0}^{j-s}
r_{2,j}^{1}f_{2,j-s-\ell}^{1}\\&+\sum_{j=s}^{N}\sum_{\ell=0}^{j-s}\sum_{p=1}^{N}\left(
                                                                               \begin{array}{c}
                                                                                 p+\ell+1 \\
                                                                                 \ell \\
                                                                               \end{array}
                                                                             \right)
r_{2,j}^{1}f_{2,j-s-\ell}^{1}\left[\frac{(-1)^{\ell}H_{p}(x,t)}{(v_{2}-v_{1})^{p+\ell}}+
\frac{(-1)^{\ell}J_{p}(x,t)}{(v_{2}+v_{1}^{*})^{p+\ell}}\right],\\
L_{s}(x,t)&=\sum_{j=s}^{N}\sum_{\ell=0}^{j-s}(-1)^{s}
r_{2,j}^{2}f_{2,j-s-\ell}^{2}\\&+\sum_{j=s}^{N}\sum_{\ell=0}^{j-s}\sum_{p=1}^{N}\left(
                                                                               \begin{array}{c}
                                                                                 p+\ell+1 \\
                                                                                 \ell \\
                                                                               \end{array}
                                                                             \right)
(-1)^{s}r_{2,j}^{2}f_{2,j-s-\ell}^{2}\left[\frac{(-1)^{\ell}H_{p}(x,t)}{(-v_{2}^{*}-v_{1})^{p+\ell}}+
\frac{(-1)^{\ell}J_{p}(x,t)}{(-v_{2}^{*}+v_{1}^{*})^{p+\ell}}\right].
\end{split}\end{align}
Let us introduce
\begin{subequations}
\begin{align}
|\eta_{1}\rangle&=\left(
                   \begin{array}{ccc}
                     \eta_{1,1} & \cdots & \eta_{1,N} \\
                   \end{array}
                 \right)^{T},\qquad\qquad~~
|\eta_{2}\rangle=\left(
                   \begin{array}{ccc}
                     \eta_{2,1} & \cdots & \eta_{2,N} \\
                   \end{array}
                 \right)^{T},\\
|\tilde{\eta}_{1}\rangle&=\left(
                   \begin{array}{ccc}
                    \tilde{ \eta}_{1,1} & \cdots & \tilde{\eta}_{1,N} \\
                   \end{array}
                 \right)^{T},\qquad\qquad~~
|\tilde{\eta}_{2}\rangle=\left(
                   \begin{array}{ccc}
                     \tilde{\eta}_{2,1} & \cdots & \tilde{\eta}_{2,N} \\
                   \end{array}
                 \right)^{T},\\
\eta_{1,s}&=\sum_{j=s}^{N}\sum_{\ell=0}^{j-s}r_{1,j}^{1}f_{1,j-s-\ell}^{1}
\frac{(-1)^{\ell}iq_{-}}{(v_{1})^{\ell+1}},\qquad~~
\tilde{\eta}_{1,s}=\sum_{j=s}^{N}\sum_{\ell=0}^{j-s}r_{2,j}^{1}f_{2,j-s-\ell}^{1},\\
\eta_{2,s}&=\sum_{j=s}^{N}\sum_{\ell=0}^{j-s}(-1)^{s}r_{1,j}^{2}f_{1,j-s-\ell}^{2}
\frac{(-1)^{\ell}iq_{-}}{(-v_{1}^{*})^{\ell+1}},~~
\tilde{\eta}_{2,s}=\sum_{j=s}^{N}\sum_{\ell=0}^{j-s}(-1)^{s}r_{2,j}^{2}f_{2,j-s-\ell}^{2},\\
\Xi_{1}&=[\Xi_{1,sp}]_{N\times N}=\sum_{j=s}^{N}\sum_{\ell=0}^{j-s}\left(
                                                                               \begin{array}{c}
                                                                                 p+\ell+1 \\
                                                                                 \ell \\
                                                                               \end{array}
                                                                             \right)
r_{1,j}^{1}f_{1,j-s-\ell}^{1}\frac{(-1)^{\ell}}{(v_{1}-v_{2})^{p+\ell}},\\
\Xi_{2}&=[\Xi_{2,sp}]_{N\times N}=\sum_{j=s}^{N}\sum_{\ell=0}^{j-s}\left(
                                                                               \begin{array}{c}
                                                                                 p+\ell+1 \\
                                                                                 \ell \\
                                                                               \end{array}
                                                                             \right)
r_{1,j}^{1}f_{1,j-s-\ell}^{1}\frac{(-1)^{\ell}}{(v_{1}+v_{2}^{*})^{p+\ell}},\\
\Xi_{3}&=[\Xi_{3,sp}]_{N\times N}=\sum_{j=s}^{N}\sum_{\ell=0}^{j-s}\sum_{p=1}^{N}\left(
                                                                               \begin{array}{c}
                                                                                 p+\ell+1 \\
                                                                                 \ell \\
                                                                               \end{array}
                                                                             \right)
(-1)^{s}r_{1,j}^{2}f_{1,j-s-\ell}^{2}\frac{(-1)^{\ell}}{(-v_{1}^{*}-v_{2})^{p+\ell}},\\
\Xi_{4}&=[\Xi_{4,sp}]_{N\times N}=\sum_{j=s}^{N}\sum_{\ell=0}^{j-s}\sum_{p=1}^{N}\left(
                                                                               \begin{array}{c}
                                                                                 p+\ell+1 \\
                                                                                 \ell \\
                                                                               \end{array}
                                                                             \right)
(-1)^{s}r_{1,j}^{2}f_{1,j-s-\ell}^{2}
\frac{(-1)^{\ell}}{(-v_{1}^{*}+v_{2}^{*})^{p+\ell}},\\
\Xi_{5}&=[\Xi_{5,sp}]_{N\times N}=\sum_{j=s}^{N}\sum_{\ell=0}^{j-s}\left(
                                                                               \begin{array}{c}
                                                                                 p+\ell+1 \\
                                                                                 \ell \\
                                                                               \end{array}
                                                                             \right)
r_{1,j}^{1}f_{1,j-s-\ell}^{1}\frac{(-1)^{\ell}}{(v_{1}-v_{2})^{p+\ell}},
\end{align}\end{subequations}
\begin{subequations}\begin{align}
\Xi_{6}&=[\Xi_{6,sp}]_{N\times N}=\sum_{j=s}^{N}\sum_{\ell=0}^{j-s}\left(
                                                                               \begin{array}{c}
                                                                                 p+\ell+1 \\
                                                                                 \ell \\
                                                                               \end{array}
                                                                             \right)
r_{1,j}^{1}f_{1,j-s-\ell}^{1}\frac{(-1)^{\ell}}{(v_{1}+v_{2}^{*})^{p+\ell}},\\
\Xi_{7}&=[\Xi_{7,sp}]_{N\times N}=\sum_{j=s}^{N}\sum_{\ell=0}^{j-s}\sum_{p=1}^{N}\left(
                                                                               \begin{array}{c}
                                                                                 p+\ell+1 \\
                                                                                 \ell \\
                                                                               \end{array}
                                                                             \right)
(-1)^{s}r_{2,j}^{2}f_{2,j-s-\ell}^{2}\frac{(-1)^{\ell}}{(-v_{2}^{*}-v_{1})^{p+\ell}},\\
\Xi_{8}&=[\Xi_{8,sp}]_{N\times N}=\sum_{j=s}^{N}\sum_{\ell=0}^{j-s}\sum_{p=1}^{N}\left(
                                                                               \begin{array}{c}
                                                                                 p+\ell+1 \\
                                                                                 \ell \\
                                                                               \end{array}
                                                                             \right)
(-1)^{s}r_{2,j}^{2}f_{2,j-s-\ell}^{2}
\frac{(-1)^{\ell}}{(-v_{2}^{*}+v_{1}^{*})^{p+\ell}},\\
|H\rangle&=\left(
             \begin{array}{ccc}
               H_{1} & \cdots & H_{N} \\
             \end{array}
           \right)^{T},\qquad\qquad
           |K\rangle=\left(
             \begin{array}{ccc}
               K_{1} & \cdots & K_{N} \\
             \end{array}
           \right)^{T},\\
|J\rangle&=\left(
             \begin{array}{ccc}
               J_{1} & \cdots & J_{N} \\
             \end{array}
           \right)^{T},\qquad\qquad~~
|L\rangle=\left(
             \begin{array}{ccc}
               L_{1} & \cdots & L_{N} \\
             \end{array}
           \right)^{T},\\
\Xi&=\left(
      \begin{array}{cc}
        \Xi_{1} & \Xi_{2} \\
        \Xi_{3} & \Xi_{4} \\
      \end{array}
    \right),~~\quad
\widetilde{\Xi}=\left(
      \begin{array}{cc}
        \Xi_{5} & \Xi_{6} \\
        \Xi_{7} & \Xi_{8} \\
      \end{array}
    \right),~~\quad
I_{\sigma}=\left(
             \begin{array}{cc}
               I & 0 \\
               0 & I \\
             \end{array}
           \right)_{2N\times2N}.
\end{align}\end{subequations}
By combining $H_{s}(x,t), K_{s}(x,t), J_{s}(x,t)$ and $L_{s}(x,t)$, we can get this system
\begin{subequations}\begin{align}
I_{\sigma}|R_{1}\rangle-\Xi|R_{2}\rangle=\alpha_{1},\\
-\widetilde{\Xi}|R_{1}\rangle+I_{\sigma}|R_{2}\rangle=\alpha_{2},
\end{align}\end{subequations}
where
\begin{subequations}\begin{align}
|R_{1}\rangle=\left(
                \begin{array}{cc}
                  |H\rangle & |J\rangle \\
                \end{array}
              \right)^{T},~~
|R_{2}\rangle=\left(
                \begin{array}{cc}
                  |K\rangle & |L\rangle \\
                \end{array}
              \right)^{T},\\
\alpha_{1}=\left(
                \begin{array}{cc}
                  |\eta_{1}\rangle & |\eta_{2}\rangle \\
                \end{array}
              \right)^{T},~~
\alpha_{2}=\left(
                \begin{array}{cc}
                  |\tilde{\eta}_{1}\rangle & |\tilde{\eta}_{2}\rangle \\
                \end{array}
              \right)^{T}.
\end{align}\end{subequations}
Through direct calculation, we have
\begin{align}
|R_{2}\rangle=(I_{\sigma}-\widetilde{\Xi}\Xi)^{-1}\alpha_{2}+(I_{\sigma}-\widetilde{\Xi}\Xi)^{-1}
\widetilde{\Xi}\alpha_{1},
\end{align}
then
\begin{align}
\begin{split}
M_{12}&=\frac{i}{z}q_{-}+\sum_{s=1}^{N}\left(\frac{K_{s}(x,t)}{(z-v_{2})^{s}}+
\frac{L_{s}(x,t)}{(z+v_{2}^{*})^{s}}\right)
=\frac{i}{z}q_{-}+\langle W|R_{2}\rangle\\
&=\frac{i}{z}q_{-}+\frac{\det \left((I_{\sigma}-\widetilde{\Xi}\Xi)+\alpha_{2}\langle W|\right)+\det\left((I_{\sigma}-\widetilde{\Xi}\Xi)+\widetilde{\Xi}\alpha_{2}\langle W|\right)} {\det(I_{\sigma}-\widetilde{\Xi}\Xi)}-2,
\end{split}
\end{align}
where
\begin{align*}
\langle W|=\left(
             \begin{array}{cc}
               \langle W_{1}| & \langle W_{2}| \\
             \end{array}
           \right),
\langle W_{1}|=\left(\frac{1}{z-v_{2}}, \cdots, \frac{1}{(z-v_{2})^{N}}\right),
\langle W_{2}|=\left(\frac{1}{z+v_{2}^{*}}, \cdots, \frac{1}{(z+v_{2}^{*})^{N}}\right).
\end{align*}
\begin{thm}\label{sol-1}
The $N$-th order soliton of the nonlocal mKdV equation with the nonzero boundary condition is
\begin{align}
q(x,t)=q_{-}-i\left[\frac{\det \left((I_{\sigma}-\widetilde{\Xi}\Xi)+\alpha_{2}\langle W_0|\right)+\det\left((I_{\sigma}-\widetilde{\Xi}\Xi)+\widetilde{\Xi}\alpha_{1}\langle W_0|\right)} {\det(I_{\sigma}-\widetilde{\Xi}\Xi)}-2\right],
\end{align}
where
\begin{align*}
\langle W_{0}|=(1,0,\cdots,0,1,0,\cdots,0)_{1\times2N}.
\end{align*}
\end{thm}

In accordance with the results of Theorem \ref{sol-1},  the expressions and propagation behaviors of one-soliton and two-soliton solutions are obtained,  respectively. When the dynamic behaviors of one-soliton solution are discussed, we only consider two cases of $\delta\sigma=1$, namely $\delta=\sigma=1$ and $\delta=\sigma=-1$. For $\delta\sigma=1$, the discrete spectrum is the set $\left\{z_n, -z_n^*, \frac{q_{0}^2}{z_n}, -\frac{q_{0}^2}{z_n^*}\right\}$ while the discrete spectrum is the set $\left\{z_n, -z_n^*, -\frac{q_{0}^2}{z_n}, \frac{q_{0}^2}{z_n^*}\right\}$ for $\delta\sigma=-1$, we can summarize the discrete spectrum in these two different cases as a characteristic of discrete spectrum, which is $\left\{v_1, -v_1^*, \frac{q_{0}^2}{v_2}, -\frac{q_{0}^2}{v_2^*}\right\}$. In \cite{PD-2019}, we know that when $\sigma=1, \delta=-1$, there is no reflectionless potential of  the defocusing nonlocal mKdV equation. Therefore, based on the above considerations, we only study the dynamic behavior of one-soliton solutions when $\delta=\sigma=1$ and $\delta=\sigma=-1$ and the dynamic behavior of two-soliton solution when $\sigma=-1, \delta=-1$.

\textbf{Case 1:} $\sigma\delta=1$. In this case, figure 1 (left) displays the distribution of the discrete spectrum. Let $z_{1}$ be the one first-order pole points of the scattering data $s_{11}(z)$, the elements $\Xi$ and $\widetilde{\Xi}$ shown in Theorem \ref{sol-1} are a $2\times2$ matrix defined by
\begin{subequations}
\begin{align}
\Xi_{11}=\frac{r_{1,1}^{1}f_{1,0}^{1}}{v_{1}-v_{2}},~
\Xi_{12}=\frac{r_{1,1}^{1}f_{1,0}^{1}}{v_{1}+v_{2}^{*}},~
\Xi_{21}=\frac{-r_{1,1}^{2}f_{1,0}^{2}}{-v_{1}^{*}-v_{2}},~
\Xi_{22}=\frac{-r_{1,1}^{2}f_{1,0}^{2}}{-v_{1}^{*}+v_{2}^{*}},\\
\widetilde{\Xi}_{11}=\frac{r_{2,1}^{1}f_{2,0}^{1}}{v_{2}-v_{1}},~
\widetilde{\Xi}_{12}=\frac{r_{2,1}^{1}f_{2,0}^{1}}{v_{2}+v_{1}^{*}},~
\widetilde{\Xi}_{21}=\frac{-r_{2,1}^{2}f_{2,0}^{2}}{-v_{2}^{*}-v_{1}},~
\widetilde{\Xi}_{22}=\frac{-r_{2,1}^{2}f_{2,0}^{2}}{-v_{2}^{*}+v_{1}^{*}},
\end{align}
\end{subequations}
the element $\eta$ and $\tilde{\eta}$ are the column vector
\begin{align*}
\eta_{1,s}=\frac{iq_{-}r_{1,1}^{1}f_{1,0}^{1}}{v_{1}},~
\eta_{2,s}\frac{iq_{-}r_{1,1}^{2}f_{1,0}^{2}}{-v_{1}^{*}},~
\tilde{\eta}_{1,s}=r_{2,1}^{1}f_{2,0}^{1},~
\tilde{\eta}_{2,s}=-r_{2,1}^{2}f_{2,0}^{2},
\end{align*}
and $\langle W_{0}|=(1, 1)$.
In accordance with Theorem \ref{sol-1}, the exact solution of nonlocal mKdV equation is derived. By selecting appropriate parameters, the following figures are obtained.
\begin{figure} [h]
 \centering
 \begin{minipage}{0.25\textwidth}
 \centering
 \includegraphics [scale=0.24]{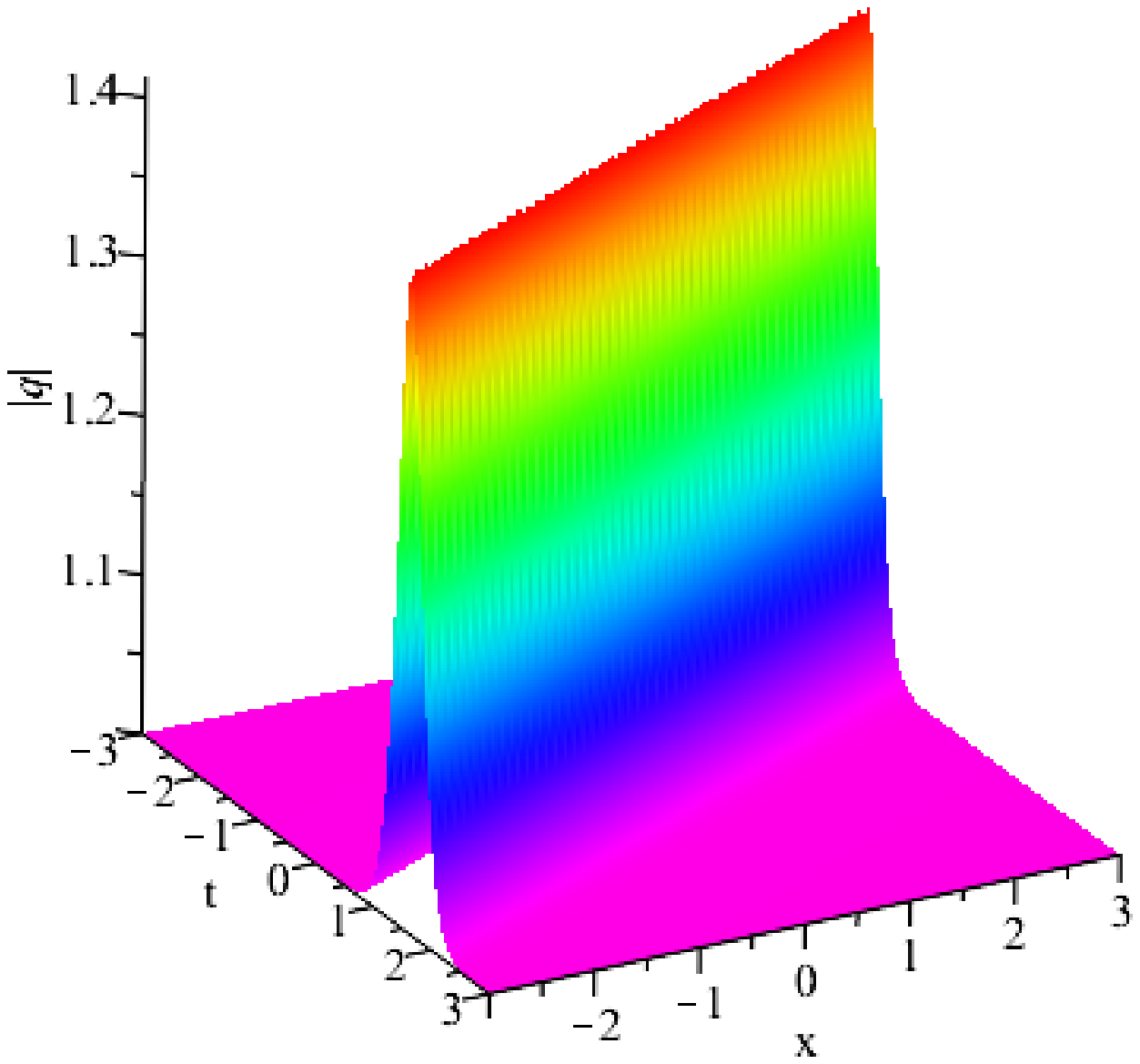}\\
 \scriptsize{($2a$)}
 \end{minipage}
 \begin{minipage}{0.35\textwidth}
 \centering
 \includegraphics [scale=0.2]{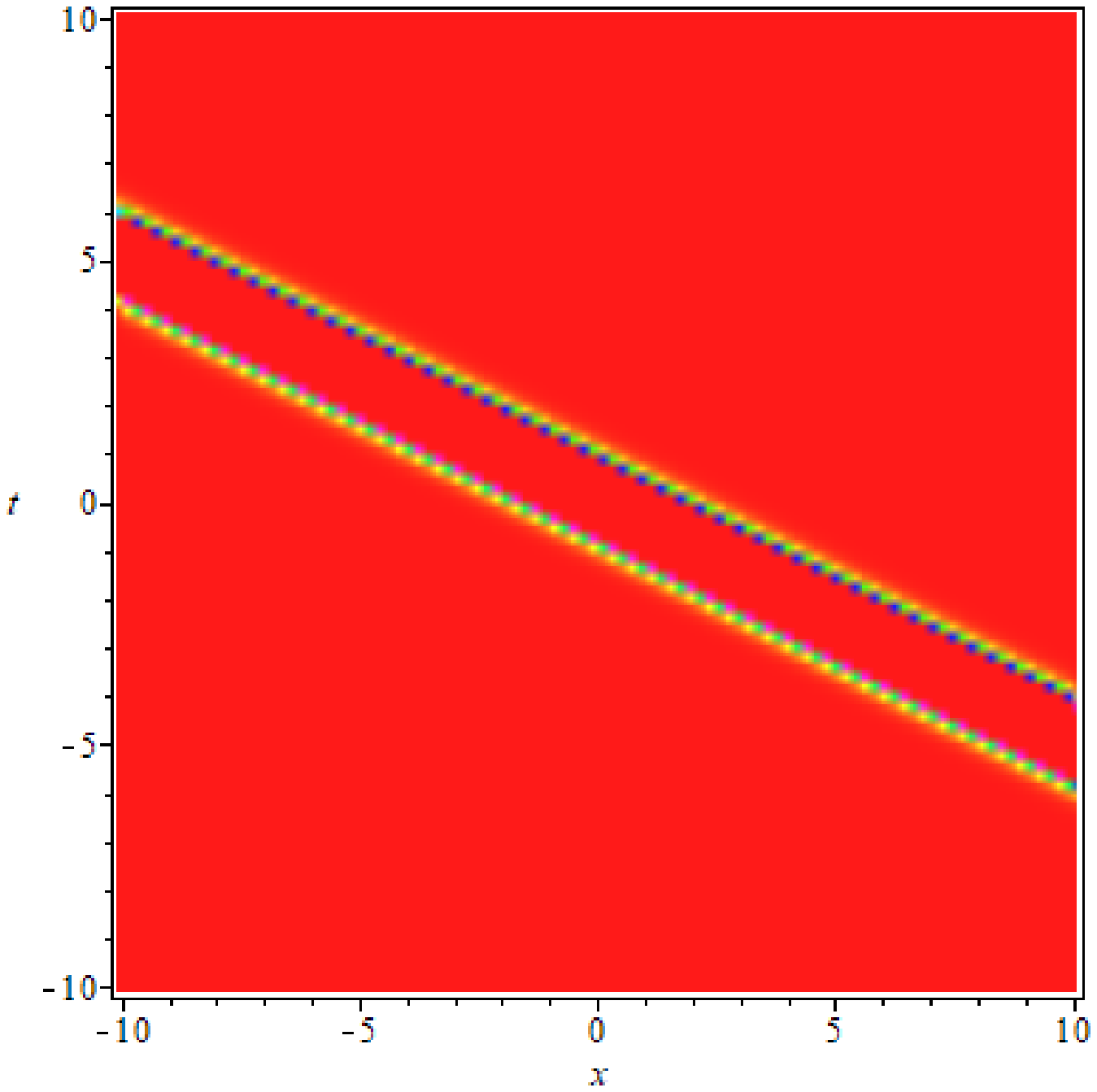}\\
 \scriptsize{($2b$)}
 \end{minipage}
 \begin{minipage}{0.32\textwidth}
 \centering
 \includegraphics [scale=0.2]{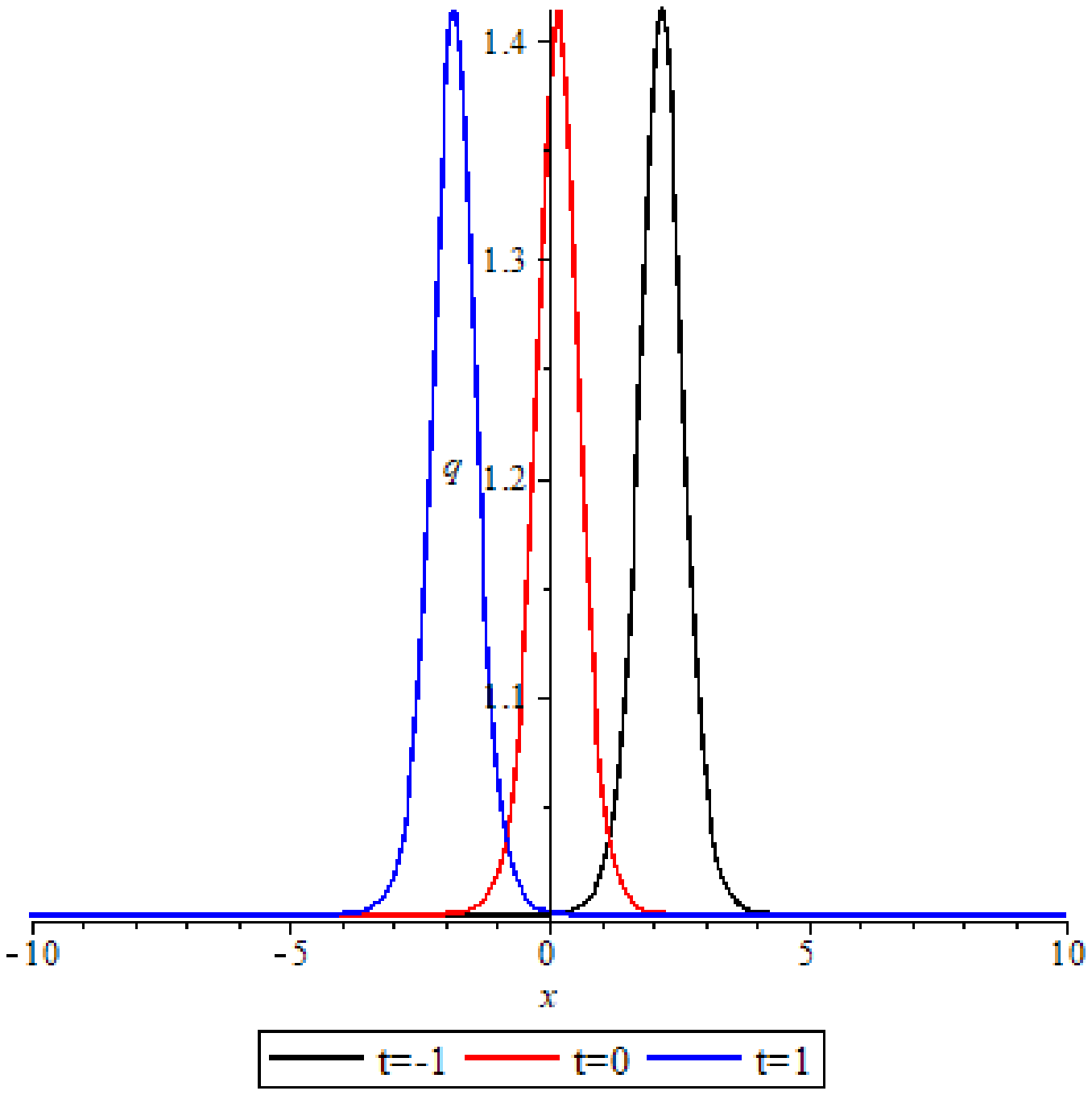}\\
 \scriptsize{($2c$)}
 \end{minipage}
\noindent { \small \textbf{Figure 2.}
The one-soliton solution  of the defocusing nonlocal cmKdV equation with $\delta=1$. Figure $(\textbf{2a})$ shows the solution graph with $v_1=i$, $r_{1,1}^{1}=1, r_{1,1}^{2}=2, r_{2,1}^{1}=3$ and $r_{2,1}^{2}=1$. Figure \textbf{(2b)} denotes the density of the solution.  Figure $(\textbf{2c})$ presents the dynamic behavior of the one-soliton solutions at different times. }
\end{figure}

\begin{figure} [h]
 \centering
 \begin{minipage}{0.25\textwidth}
 \centering
 \includegraphics [scale=0.22]{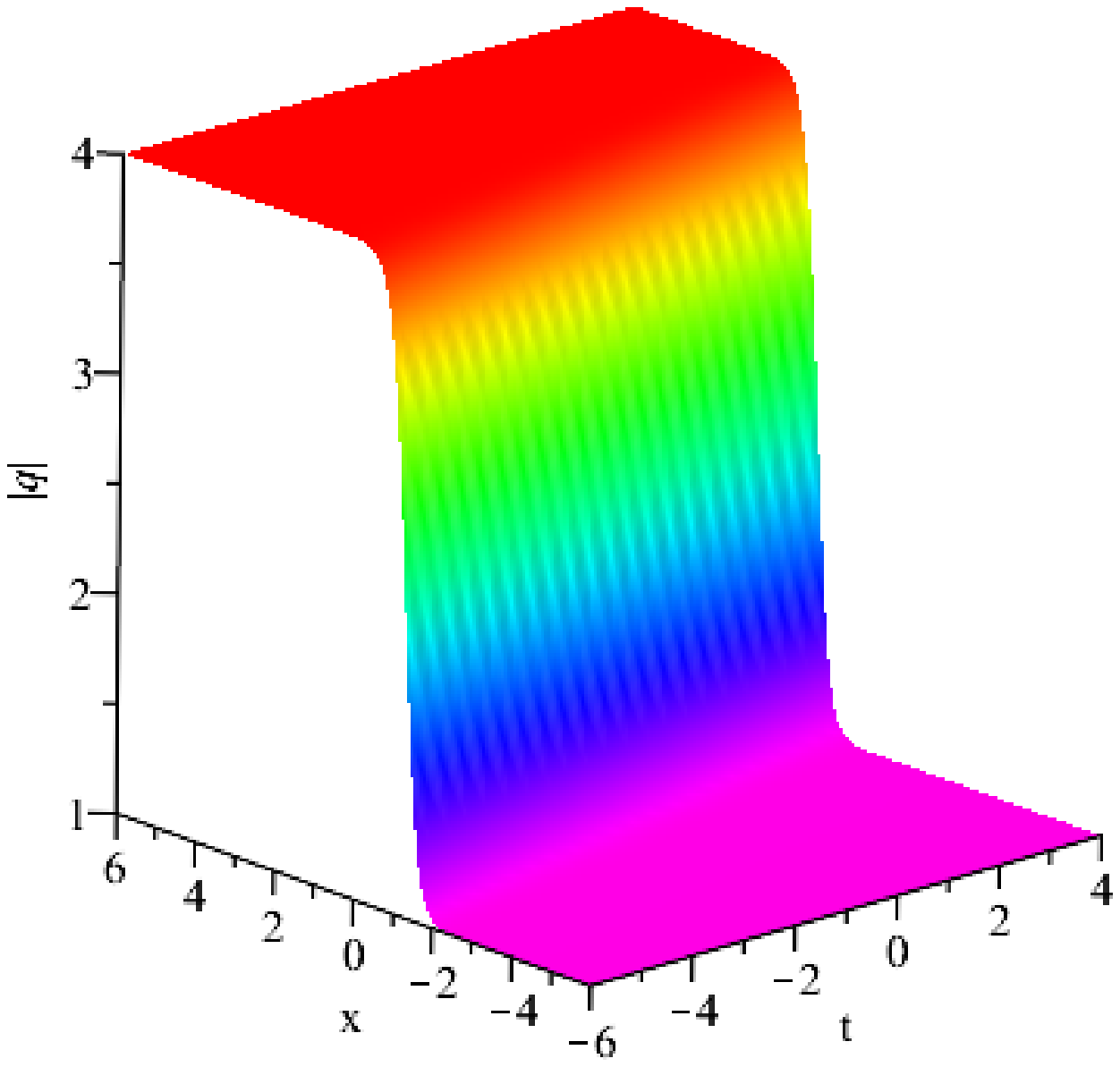}\\
 \scriptsize{($3a$)}
 \end{minipage}
 \begin{minipage}{0.35\textwidth}
 \centering
 \includegraphics [scale=0.34]{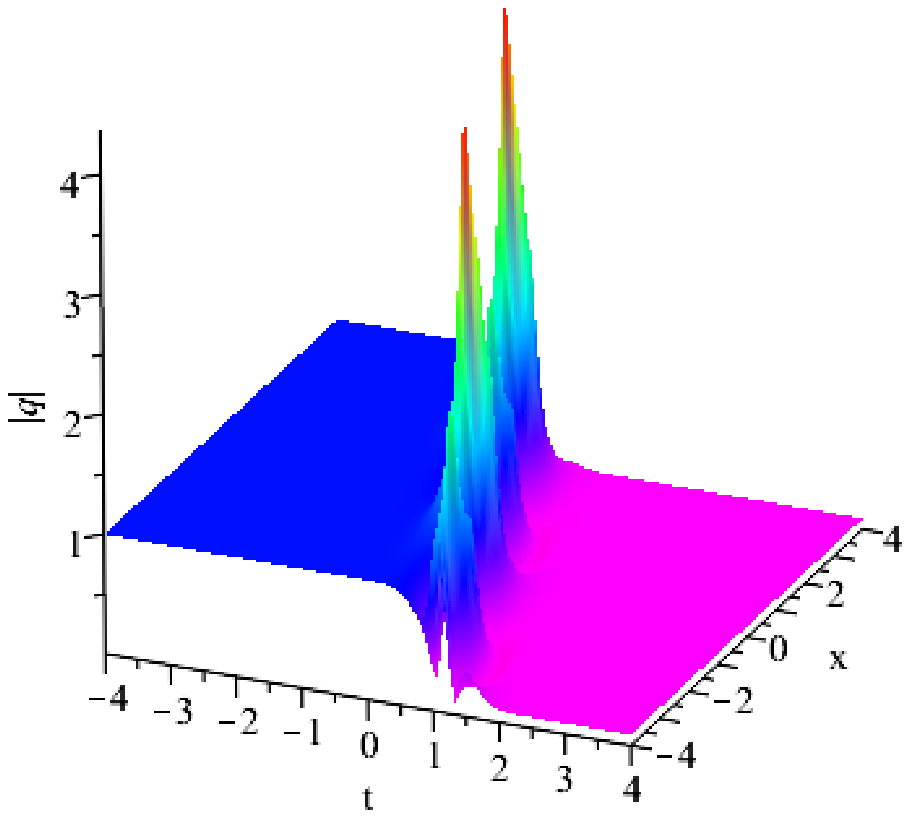}\\
 \scriptsize{($3b$)}
 \end{minipage}
 \begin{minipage}{0.32\textwidth}
 \centering
 \includegraphics [scale=0.34]{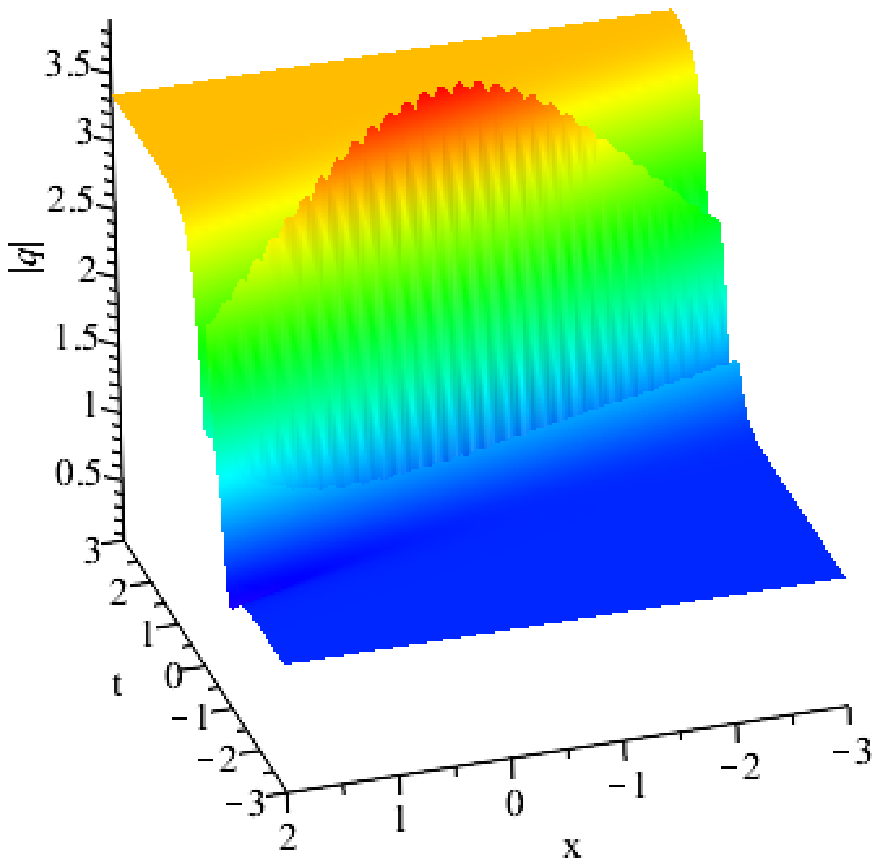}\\
 \scriptsize{($3c$)}
 \end{minipage}
\noindent { \small \textbf{Figure 3.}
The one-soliton solution  of the focusing nonlocal cmKdV equation with $\delta=-1$. Figure $(\textbf{3a})$ shows the solution graph with $v_1=\frac{1}{2}i$, $r_{1,1}^{1}=1, r_{1,1}^{2}=2, r_{2,1}^{1}=3$ and $r_{2,1}^{2}=1$. Figure \textbf{(3b)} shows the solution graph with $v_1=1+2i$. Figure $(\textbf{3c})$ shows the solution graph. }
\end{figure}

Fig. 2 and Fig. 3 show the propagation behaviors of one-soliton solution in the case of $\delta=\sigma=1$ and $\delta=\sigma=-1$. Figure 2a illustrates the anti-dark solution, which can be intuitively observed as a bounded solution. Figure 3a displays the kink solution when the selected parameter is pure imaginary number, which is different from figure 2a because $q_{+}=\delta q_-$ in nonzero boundary condition $\mathop{\lim}\limits_{x\rightarrow\pm\infty}q(x,t)=q_{\pm}$.  When the selected parameter is a non-pure imaginary number, figures 3b and 3c are both kink-type solutions and there will be breathing solutions near $t=0$. When $q_{-}=-q_+$, it can be intuitively observed from the figures that the amplitude on the left side of $t=0$ is significantly higher than the amplitude on the right side.

\textbf{Case 2:} $\delta=-1, \sigma=-1$. In this case, figure 1 (right) displays the distribution of the discrete spectrum. Let $z_{1}$ be the two first-order pole points of the scattering data $s_{11}(z)$, the elements $\Xi$ and $\widetilde{\Xi}$ shown in Theorem \ref{sol-1} are a $2\times2$ matrix defined by
\begin{figure} [h]
 \centering
 \begin{minipage}{0.25\textwidth}
 \centering
 \includegraphics [scale=0.22]{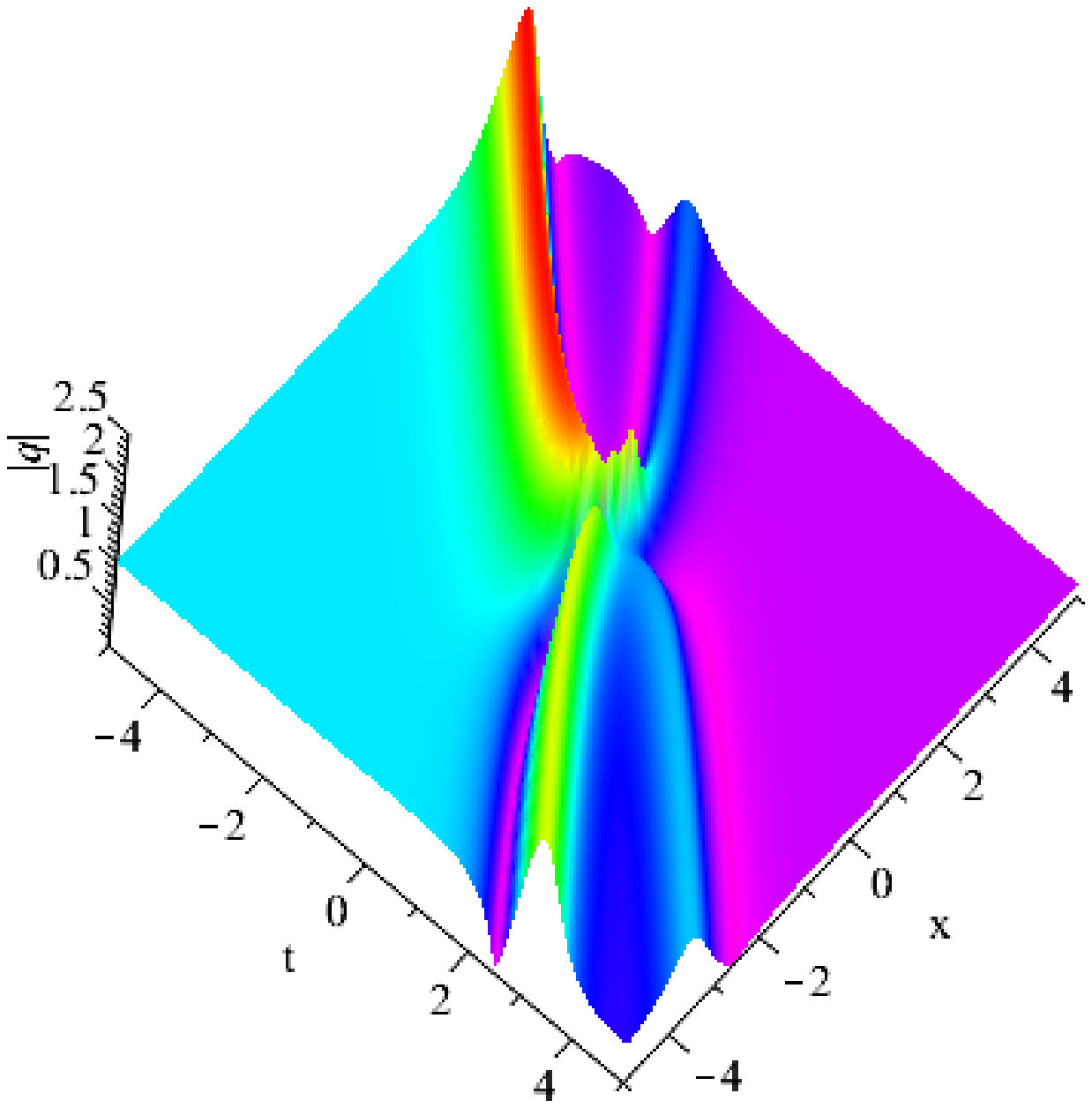}\\
 \scriptsize{($4a$)}
 \end{minipage}
 \begin{minipage}{0.35\textwidth}
 \centering
 \includegraphics [scale=0.18]{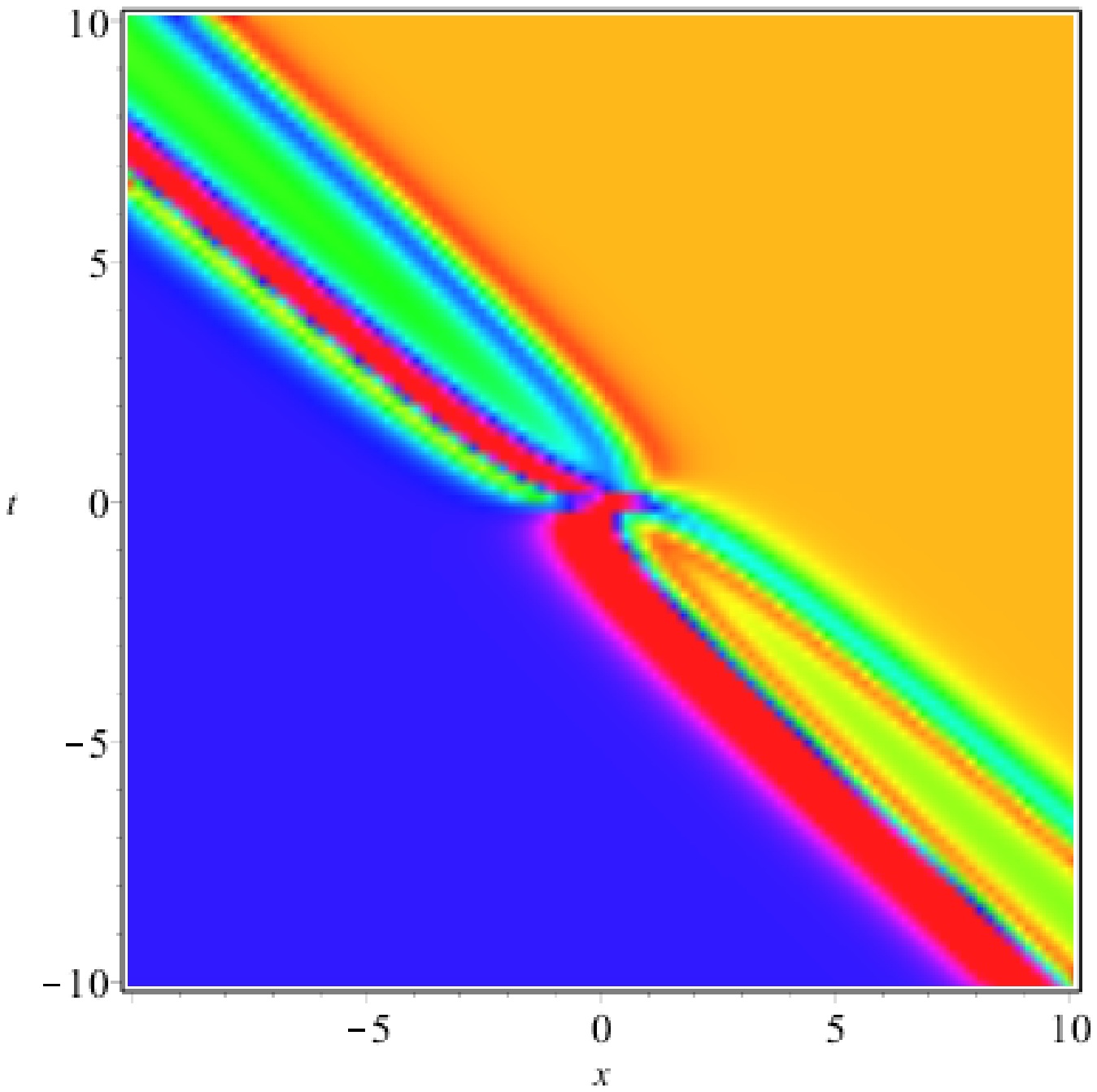}\\
 \scriptsize{($4b$)}
 \end{minipage}
 \begin{minipage}{0.30\textwidth}
 \centering
 \includegraphics [scale=0.18]{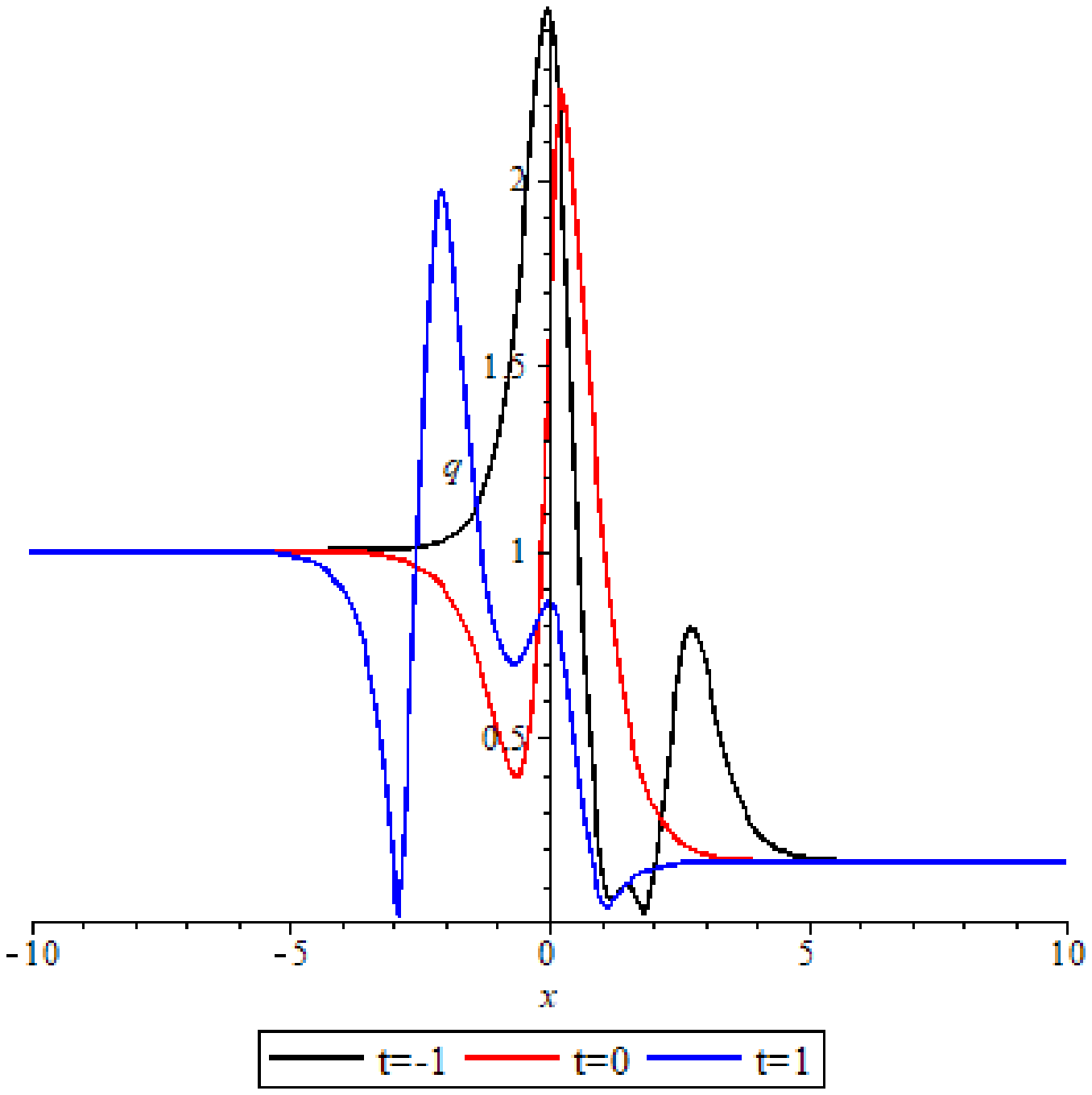}\\
 \scriptsize{($4c$)}
 \end{minipage}
 \centering
 \begin{minipage}{0.25\textwidth}
 \centering
 \includegraphics [scale=0.22]{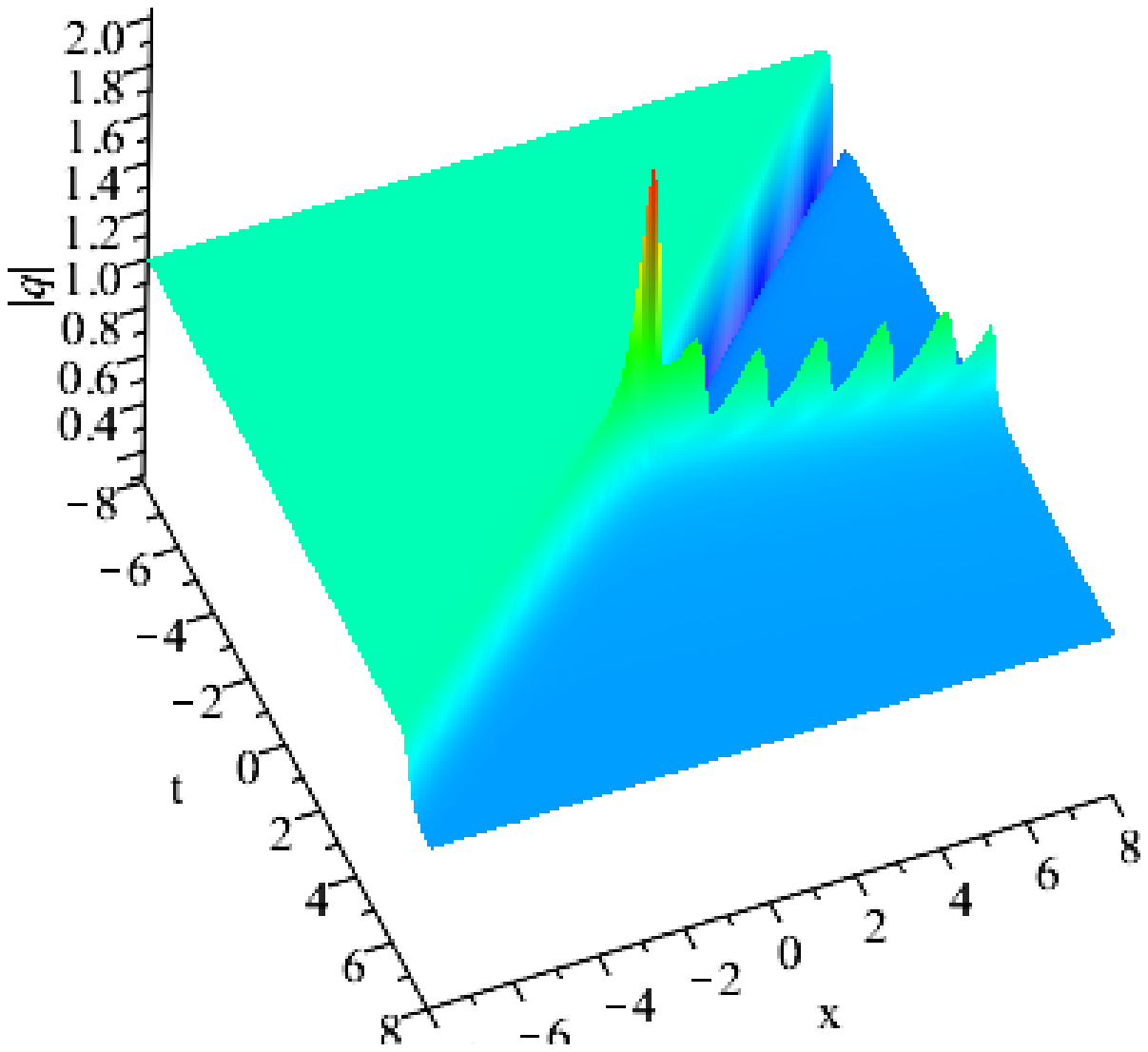}\\
 \scriptsize{($4d$)}
 \end{minipage}
 \begin{minipage}{0.35\textwidth}
 \centering
 \includegraphics [scale=0.18]{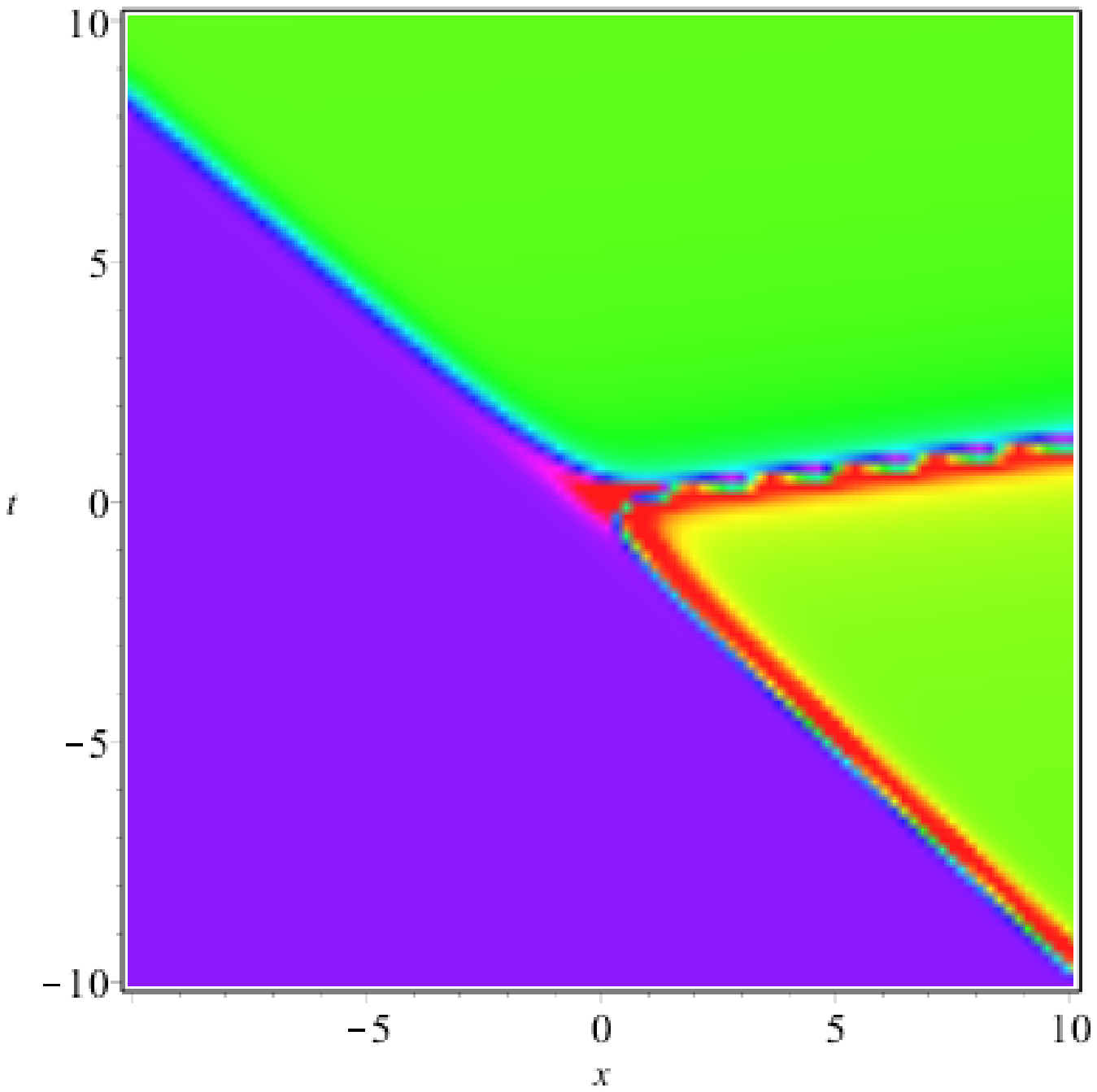}\\
 \scriptsize{($4e$)}
  \end{minipage}
 \begin{minipage}{0.3\textwidth}
 \centering
 \includegraphics [scale=0.18]{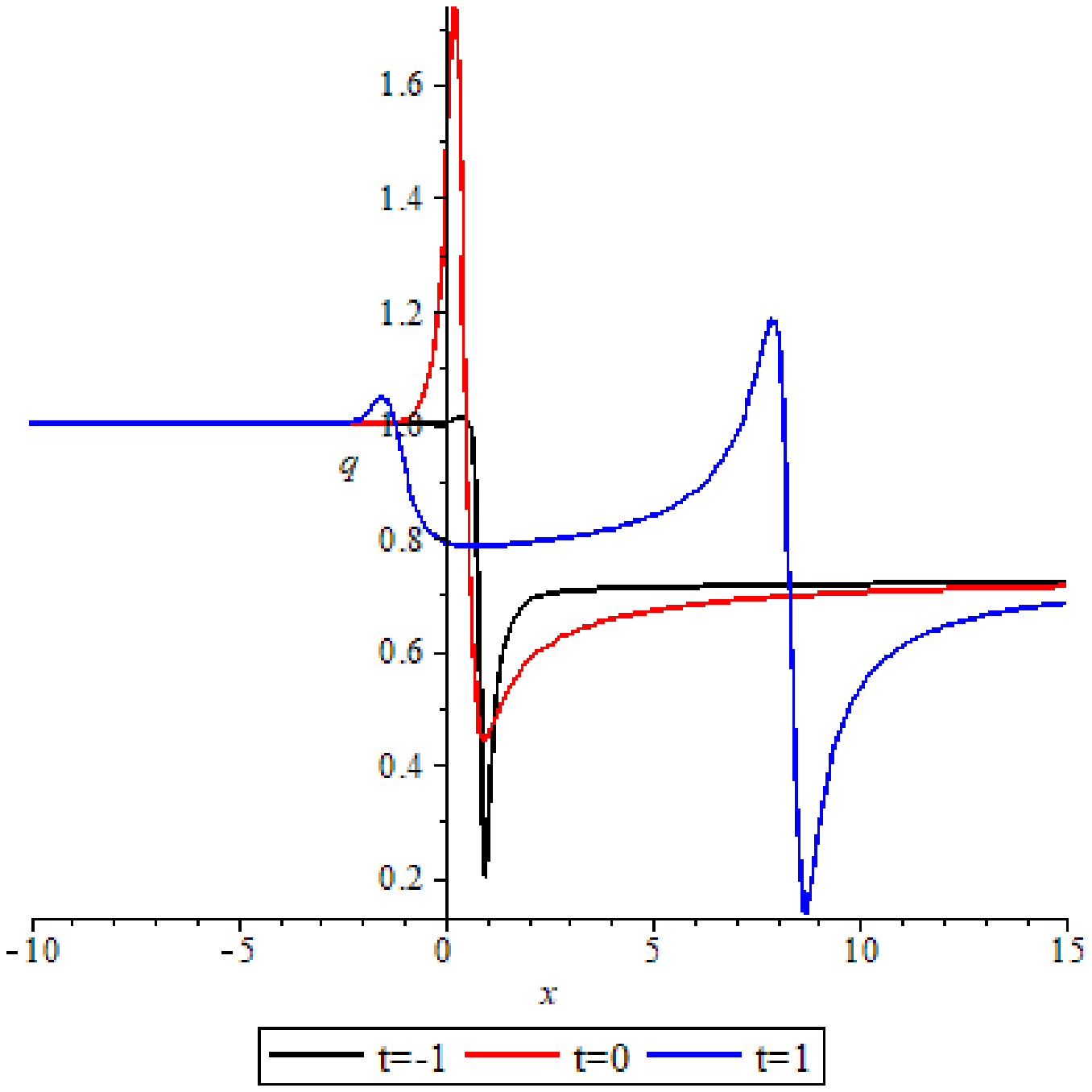}\\
 \scriptsize{($4f$)}
 \end{minipage}

\noindent { \small \textbf{Figure 4.}
The two-soliton solution  of the nonlocal focusing cmKdV equation with $\delta=-1$. Figure $(\textbf{4a})$ shows the solution graph with $v_1=\frac{\pi}{2}i$, $r_{1,1}^{1}=1, r_{1,1}^{2}=3, r_{2,1}^{1}=1, r_{2,1}^{2}=2, r_{1,2}^{1}=1,$ $r_{1,2}^{2}=3, r_{2,2}^{1}=1$ and $r_{2,2}^{2}=4$. Figure \textbf{(4b)} denotes the density of the solution. Figure $(\textbf{4c})$ represents the dynamic behavior of the two-soliton solutions at different times. Figure $(\textbf{4d})$ shows the solution graph with $v_1=\frac{\pi}{2}i$, $r_{1,1}^{1}=1, r_{1,1}^{2}=3, r_{2,1}^{1}=1,r_{2,1}^{2}=2, r_{1,2}^{1}=1, r_{1,2}^{2}=3,$ $ r_{2,2}^{1}=1$ and $r_{2,2}^{2}=1$. Figure \textbf{(4e)} denotes the density of the solution. Figure $(\textbf{4f})$ represents the dynamic behavior of the two-soliton solutions at different times.}
\end{figure}
\begin{align*}
[\Xi_{11}]_{11}&=\frac{r_{1,1}^{1}f_{1,0}^{1}}{v_1-v_2}+\frac{r_{1,2}^{1}f_{1,1}^{1}}{v_1-v_2}-
\frac{r_{1,2}^{1}f_{1,0}^{1}}{(v_1-v_2)^{2}},\qquad\qquad\quad~~
[\Xi_{11}]_{21}=\frac{r_{1,2}^{1}f_{1,0}^{1}}{v_1-v_2},\\
[\Xi_{11}]_{12}&=\frac{r_{1,1}^{1}f_{1,0}^{1}}{(v_1-v_2)^{2}}+\frac{r_{1,2}^{1}f_{1,1}^{1}}{(v_1-v_2)^{2}}-
\frac{2r_{1,2}^{1}f_{1,0}^{1}}{(v_1-v_2)^{3}},\qquad\quad~~
[\Xi_{11}]_{22}=\frac{r_{1,2}^{1}f_{1,0}^{1}}{(v_1-v_2)^{2}},\\
[\Xi_{12}]_{11}&=\frac{r_{1,1}^{1}f_{1,0}^{1}}{v_1+v_2^*}+\frac{r_{1,2}^{1}f_{1,1}^{1}}{v_1+v_2^*}-
\frac{r_{1,2}^{1}f_{1,0}^{1}}{(v_1+v_2^*)^{2}},\qquad\qquad\quad~~~~
[\Xi_{12}]_{21}=\frac{r_{1,2}^{1}f_{1,0}^{1}}{v_1+v_2^*},\\
[\Xi_{12}]_{12}&=\frac{r_{1,1}^{1}f_{1,0}^{1}}{(v_1+v_2^*)^{2}}+\frac{r_{1,2}^{1}f_{1,1}^{1}}{(v_1+v_2^*)^{2}}-
\frac{2r_{1,2}^{1}f_{1,0}^{1}}{(v_1+v_2^*)^{3}},\qquad\qquad
[\Xi_{12}]_{22}=\frac{r_{1,2}^{1}f_{1,0}^{1}}{(v_1+v_2^*)^{2}},\\
[\Xi_{21}]_{11}&=-\frac{r_{1,1}^{2}f_{1,0}^{2}}{-v_1^*-v_2}-\frac{r_{1,2}^{2}f_{1,1}^{2}}{-v_1^*-v_2}+
\frac{r_{1,2}^{2}f_{1,0}^{2}}{(-v_1^*-v_2)^{2}},\qquad\qquad
[\Xi_{21}]_{21}=\frac{r_{1,2}^{2}f_{1,0}^{2}}{-v_1^*-v_2},\\
[\Xi_{21}]_{12}&=-\frac{r_{1,1}^{2}f_{1,0}^{2}}{(-v_1^*-v_2)^{2}}-\frac{r_{1,2}^{2}f_{1,1}^{2}}{(-v_1^*-v_2)^{2}}+
\frac{2r_{1,2}^{2}f_{1,0}^{2}}{(-v_1^*-v_2)^{3}},\qquad
[\Xi_{21}]_{22}=\frac{r_{1,2}^{2}f_{1,0}^{2}}{(-v_1^*-v_2)^{2}},\\
[\Xi_{22}]_{11}&=-\frac{r_{1,1}^{2}f_{1,0}^{2}}{-v_1^*+v_2^*}-\frac{r_{1,2}^{2}f_{1,1}^{2}}{-v_1^*+v_2^*}+
\frac{r_{1,2}^{2}f_{1,0}^{2}}{(-v_1^*+v_2^*)^{2}},\qquad\qquad
[\Xi_{22}]_{21}=\frac{r_{1,2}^{2}f_{1,0}^{2}}{-v_1^*+v_2^*},\\
[\Xi_{22}]_{12}&=-\frac{r_{1,1}^{2}f_{1,0}^{2}}{(-v_1^*+v_2^*)^{2}}-\frac{r_{1,2}^{2}f_{1,1}^{2}}{(-v_1^*+v_2^*)^{2}}+
\frac{2r_{1,2}^{2}f_{1,0}^{2}}{(-v_1^*+v_2^*)^{3}},\qquad
[\Xi_{22}]_{22}=\frac{r_{1,2}^{2}f_{1,0}^{2}}{(-v_1^*+v_2^*)^{2}},
\end{align*}
\begin{align*}
[\widetilde{\Xi}_{11}]_{11}&=\frac{r_{2,1}^{1}f_{2,0}^{1}}{v_2-v_1}+\frac{r_{2,2}^{1}f_{2,1}^{1}}{v_2-v_1}-
\frac{r_{2,2}^{1}f_{2,0}^{1}}{(v_2-v_1)^{2}},\qquad\qquad\qquad~~~
[\widetilde{\Xi}_{11}]_{21}=\frac{r_{2,2}^{1}f_{2,0}^{1}}{v_2-v_1},\\
[\widetilde{\Xi}_{11}]_{12}&=\frac{r_{2,1}^{1}f_{2,0}^{1}}{(v_2-v_1)^{2}}+\frac{r_{2,2}^{1}f_{2,1}^{1}}{(v_2-v_1)^{2}}-
\frac{2r_{2,2}^{1}f_{2,0}^{1}}{(v_2-v_1)^{3}},\qquad\qquad~~
[\widetilde{\Xi}_{11}]_{22}=\frac{r_{2,2}^{1}f_{2,0}^{1}}{(v_2-v_1)^{2}},\\
[\widetilde{\Xi}_{12}]_{11}&=\frac{r_{2,1}^{1}f_{2,0}^{1}}{v_2+v_1^*}+\frac{r_{2,2}^{1}f_{2,1}^{1}}{v_2+v_1^*}-
\frac{r_{2,2}^{1}f_{2,0}^{1}}{(v_2+v_1^*)^{2}},\qquad\qquad\qquad~~~
[\widetilde{\Xi}_{12}]_{21}=\frac{r_{2,2}^{1}f_{2,0}^{1}}{v_2+v_1^*},\\
[\widetilde{\Xi}_{12}]_{12}&=\frac{r_{2,1}^{1}f_{2,0}^{1}}{(v_2+v_1^*)^{2}}+\frac{r_{2,2}^{1}f_{2,1}^{1}}{(v_2+v_1^*)^{2}}-
\frac{2r_{1,2}^{1}f_{1,0}^{1}}{(v_1+v_2^*)^{3}},\qquad\qquad~~
[\widetilde{\Xi}_{12}]_{22}=\frac{r_{1,2}^{1}f_{1,0}^{1}}{(v_1+v_2^*)^{2}},\\
[\widetilde{\Xi}_{21}]_{11}&=-\frac{r_{2,1}^{2}f_{2,0}^{2}}{-v_2^*-v_1}-\frac{r_{2,2}^{2}f_{2,1}^{2}}{-v_2^*-v_1}+
\frac{r_{1,2}^{2}f_{1,0}^{2}}{(-v_2^*-v_1)^{2}},\qquad\qquad
[\widetilde{\Xi}_{21}]_{21}=\frac{r_{2,2}^{2}f_{2,0}^{2}}{-v_2^*-v_1},\\
[\widetilde{\Xi}_{21}]_{12}&=-\frac{r_{2,1}^{2}f_{2,0}^{2}}{(-v_2^*-v_1)^{2}}-\frac{r_{2,2}^{2}f_{2,1}^{2}}{(-v_2^*-v_1)^{2}}+
\frac{2r_{2,2}^{2}f_{2,0}^{2}}{(-v_2^*-v_1)^{3}},\quad~~~
[\widetilde{\Xi}_{21}]_{22}=\frac{r_{2,2}^{2}f_{2,0}^{2}}{(-v_2^*-v_1)^{2}},\\
[\widetilde{\Xi}_{22}]_{11}&=-\frac{r_{2,1}^{2}f_{2,0}^{2}}{-v_2^*+v_1^*}-\frac{r_{2,2}^{2}f_{2,1}^{2}}{-v_2^*+v_1^*}+
\frac{r_{2,2}^{2}f_{2,0}^{2}}{(-v_2^*+v_1^*)^{2}},\qquad\qquad
[\widetilde{\Xi}_{22}]_{21}=\frac{r_{2,2}^{2}f_{2,0}^{2}}{-v_2^*+v_1^*},\\
[\widetilde{\Xi}_{22}]_{12}&=-\frac{r_{2,1}^{2}f_{2,0}^{2}}{(-v_2^*+v_1^*)^{2}}-\frac{r_{2,2}^{2}f_{2,1}^{2}}{(-v_2^*+v_1^*)^{2}}+
\frac{2r_{2,2}^{2}f_{2,0}^{2}}{(-v_2^*+v_1^*)^{3}},\qquad
[\widetilde{\Xi}_{22}]_{22}=\frac{r_{2,2}^{2}f_{2,0}^{2}}{(-v_2^*+v_1^*)^{2}},
\end{align*}
the element $|\eta\rangle=(\eta_{1,1},~\eta_{1,2})^{T}$ and $|\tilde{\tilde{\eta}}\rangle=(\tilde{\eta}_{1,1},~\tilde{\eta}_{1,2})^{T}$ are the column vector
\begin{align*}
\eta_{1,1}&=\frac{r_{1,1}^{1}f_{1,0}^{1}iq_-}{v_1}+\frac{r_{1,2}^{1}f_{1,1}^{1}iq_-}{v_1}-
\frac{r_{1,2}^{1}f_{1,0}^{1}iq_-}{(v_1)^2},\quad
\eta_{1,2}=\frac{r_{1,2}^{1}f_{1,0}^{1}iq_-}{v_1},\\
\eta_{2,1}&=-\frac{r_{1,1}^{2}f_{1,0}^{2}iq_-}{-v_1^*}-\frac{r_{1,2}^{2}f_{1,1}^{2}iq_-}{-v_1^*}-
\frac{r_{1,2}^{2}f_{1,0}^{2}iq_-}{(-v_1^*)^2},\quad
\eta_{2,2}=\frac{r_{1,2}^{2}f_{1,0}^{2}iq_-}{-v_1^*},\\
\tilde{\eta}_{1,1}&=r_{2,1}^{1}f_{2,0}^{1}+r_{2,2}^{1}f_{2,1}^{1},\qquad
\tilde{\eta}_{1,2}=r_{2,2}^{1}f_{2,0}^{1},\\
\tilde{\eta}_{2,1}&=-r_{2,1}^{2}f_{2,0}^{2}-r_{2,2}^{2}f_{2,1}^{2},\quad~
\tilde{\eta}_{2,2}=r_{2,2}^{2}f_{2,0}^{2},
\end{align*}
and $\langle W_{0}|=(1, 0, 1, 0)$.
According to Theorem \ref{sol-1}, the exact solution of nonlocal mKdV equation is derived. By selecting appropriate parameters, the figures are dispicted in Fig. 4.

\section{Multiple high-order pole solutions}

The general situation will be considered that  $s_{11}(z)$ has $N$ high-order zero points $z_{k}\in D^{+}$ with $k=1, 2, \ldots, N$, their powers are $n_{1}$,  $n_{2}$, \ldots, $n_{N}$, respectively. In this case, we only consider the focusing nonlocal mKdV equation. Similar to the case of a higher-order pole discussed above, by applying the Laurent series expansion, $r^{k}(z)$ can by expanded as
\begin{subequations}
\begin{align}
r^{k}(z)&=r_{0}^{1,k}(z)+\sum_{m=1}^{N}\frac{r_{1,m}^{1,k}}{(z-v_{1}^{k})^{m}},\quad
r^{k}(z)=r_{0}^{2,k}(z)+\sum_{m=1}^{N}\frac{(-1)^{m+1}r_{1,m}^{2,k}}{(z+v_{1}^{k*})^{m}},\\
\widetilde{r^{k}}(z)&=\widetilde{r_{0}^{1,k}}(z)+\sum_{m=1}^{N}\frac{r_{2,m}^{1,k}}{(z-v_{1}^{k})^{m}},\quad
\widetilde{r^{k}}(z)=\widetilde{r_{0}^{2,k}}(z)+\sum_{m=1}^{N}\frac{(-1)^{m+1}r_{2,m}^{2,k}}{(z+v_{1}^{k*})^{m}},
\end{align}
\end{subequations}
where $r_{m}$ can be written as
\begin{align*}
r_{j,m}=\lim_{z\rightarrow v_{j}^{k}}\frac{1}{(n_{k}-m)!}\frac{\partial^{n_{k}-m}}{\partial
k^{n_{k}-m}}\left[(z-v_{j}^{k})^{n_{k}}r_{j}(z)\right].
\end{align*}
The multiple solitons of the nonlocal mKdV equation are obtained via the similar way as above.
Let us introduce
\begin{subequations}
\begin{align}
|\Gamma_{1}\rangle=&\left(
                   \begin{array}{ccc}
                     \Gamma_{1}^{1} & \cdots & \Gamma_{1}^{N} \\
                   \end{array}
                 \right)^{T},\quad\quad
|\eta_{1}^{k}\rangle=\left(
                       \begin{array}{cccc}
                         |\eta_{1,1}^{k}\rangle & |\eta_{1,2}^{k}\rangle & \cdots & |\eta_{1,N}^{k}\rangle \\
                       \end{array}
                     \right)^{T},\\
|\Gamma_{1}^{k}\rangle=&\left(
                          \begin{array}{cc}
                            |\eta_{1}^{k}\rangle & |\eta_{2}^{k}\rangle \\
                          \end{array}
                        \right)^{T},~~\qquad\quad
|\eta_{2}^{k}\rangle=\left(
                       \begin{array}{cccc}
                         |\eta_{2,1}^{k}\rangle & |\eta_{2,2}^{k}\rangle & \cdots & |\eta_{2,N}^{k}\rangle \\
                       \end{array}
                     \right)^{T},\\
|\Gamma_{2}\rangle=&\left(
                   \begin{array}{ccc}
                     \Gamma_{2}^{1} & \cdots & \Gamma_{2}^{N} \\
                   \end{array}
                 \right)^{T},\qquad
|\tilde{\eta}_{1}^{k}\rangle=\left(
                       \begin{array}{cccc}
                         |\tilde{\eta}_{1,1}^{k}\rangle & |\tilde{\eta}_{1,2}^{k}\rangle & \cdots & |\tilde{\eta}_{1,N}^{k}\rangle \\
                       \end{array}
                     \right)^{T},\\
|\Gamma_{2}^{k}\rangle=&\left(
                          \begin{array}{cc}
                            |\tilde{\eta}_{1}^{k}\rangle & |\tilde{\eta}_{2}^{k}\rangle \\
                          \end{array}
                        \right)^{T},~~\qquad\quad
|\tilde{\eta}_{2}^{k}\rangle=\left(
                       \begin{array}{cccc}
                         |\tilde{\eta}_{2,1}^{k}\rangle & |\tilde{\eta}_{2,2}^{k}\rangle & \cdots & |\tilde{\eta}_{2,N}^{k}\rangle \\
                       \end{array}
                     \right)^{T},\\
\eta_{1,s}^{k}=&\sum_{j=s}^{n_{k}}\sum_{\ell=0}^{j-s}r_{1,j}^{1,k}f_{1,j-s-\ell}^{1,k}
\frac{(-1)^{\ell}iq_{-}}{(v_{1}^k)^{\ell+1}},\qquad~~
\tilde{\eta}_{1,s}^{k}=\sum_{j=s}^{n_{k}}r_{2,j}^{1,k}f_{2,j-s-\ell}^{1,k},\\
\eta_{2,s}^{k}=&\sum_{j=s}^{n_{k}}\sum_{\ell=0}^{j-s}r_{1,j}^{2,k}f_{1,j-s-\ell}^{2,k}
\frac{(-1)^{s+\ell}iq_{-}}{(-v_{1}^{k*})^{\ell+1}},\quad~~~
\tilde{\eta}_{2,s}^{k}=\sum_{j=s}^{n_{k}}(-1)^{s}r_{2,j}^{1,k}f_{2,j-s-\ell}^{1,k},\\
\Xi_{j\ell,pq}^{(1)}=&\sum_{j=p}^{n_{j}}\sum_{s_{j}=0}^{j-p}\left(
                                                                               \begin{array}{c}
                                                                                 q+s_{j}+1 \\
                                                                                 s_{j} \\
                                                                               \end{array}
                                                                             \right)
\frac{(-1)^{s_{j}}r_{1,j}^{1,j}f_{1,j-s_{j}-p}^{1,j}}{(v_{1}^{j}-v_{2}^{j})^{s_{j}+q}},\\
\Xi_{j\ell,pq}^{(2)}=&\sum_{j=p}^{n_{j}}\sum_{s_{j}=0}^{j-p}\left(
                                                                               \begin{array}{c}
                                                                                 q+s_{j}+1 \\
                                                                                 s_{j} \\
                                                                               \end{array}
                                                                             \right)
\frac{(-1)^{s_{j}}r_{1,j}^{1,j}f_{1,j-s_{j}-p}^{1,j}}{(v_{1}^{j}+v_{2}^{j*})^{s_{j}+q}},\\
\Xi_{j\ell,pq}^{(3)}=&\sum_{j=p}^{n_{j}}\sum_{s_{j}=0}^{j-p}\left(
                                                                               \begin{array}{c}
                                                                                 q+s_{j}+1 \\
                                                                                 s_{j} \\
                                                                               \end{array}
                                                                             \right)
\frac{(-1)^{s_{j}+p}r_{1,j}^{2,j}f_{1,j-s_{j}-p}^{2,j}}{(-v_{1}^{j*}-v_{2}^{j})^{s_{j}+q}},\\
\Xi_{j\ell,pq}^{(4)}=&\sum_{j=p}^{n_{j}}\sum_{s_{j}=0}^{j-p}\left(
                                                                               \begin{array}{c}
                                                                                 q+s_{j}+1 \\
                                                                                 s_{j} \\
                                                                               \end{array}
                                                                             \right)
\frac{(-1)^{s_{j}+p}r_{1,j}^{2,j}f_{1,j-s_{j}-p}^{2,j}}{(-v_{1}^{j*}+v_{2}^{j*})^{s_{j}+q}},\\
\Xi_{j\ell,pq}^{(5)}=&\sum_{j=p}^{n_{j}}\sum_{s_{j}=0}^{j-p}\left(
                                                                               \begin{array}{c}
                                                                                 q+s_{j}+1 \\
                                                                                 s_{j} \\
                                                                               \end{array}
                                                                             \right)
\frac{r_{2,j}^{1,j}f_{2,j-s_{j}-p}^{1,j}}{(v_{2}^{j}-v_{1}^{j})^{s_{j}+q}},\\
\Xi_{j\ell,pq}^{(6)}=&\sum_{j=p}^{n_{j}}\sum_{s_{j}=0}^{j-p}\left(
                                                                               \begin{array}{c}
                                                                                 q+s_{j}+1 \\
                                                                                 s_{j} \\
                                                                               \end{array}
                                                                             \right)
\frac{r_{2,j}^{1,j}f_{2,j-s_{j}-p}^{1,j}}{(v_{2}^{j}+v_{1}^{j+})^{s_{j}+q}},
\end{align}\end{subequations}
\begin{subequations}
\begin{align}
\Xi_{j\ell,pq}^{(7)}&=\sum_{j=p}^{n_{j}}\sum_{s_{j}=0}^{j-p}\left(
                                                                               \begin{array}{c}
                                                                                 q+s_{j}+1 \\
                                                                                 s_{j} \\
                                                                               \end{array}
                                                                             \right)
\frac{(-1)^{s_{j}+p}r_{2,j}^{2,j}f_{2,j-s_{j}-p}^{2,j}}{(-v_{2}^{j*}-v_{1}^{j})^{s_{j}+q}},\\
\Xi_{j\ell,pq}^{(8)}&=\sum_{j=p}^{n_{j}}\sum_{s_{j}=0}^{j-p}\left(
                                                                               \begin{array}{c}
                                                                                 q+s_{j}+1 \\
                                                                                 s_{j} \\
                                                                               \end{array}
                                                                             \right)
\frac{(-1)^{s_{j}+p}r_{1,j}^{1,j}f_{1,j-s_{j}-p}^{1,j}}{(-v_{2}^{j*}+v_{1}^{j*})^{s_{j}+q}},\\
\Omega=&\left(
          \begin{array}{cccc}
            \omega_{11} & \omega_{12} & \cdots & \omega_{1N} \\
            \omega_{21} & \omega_{22} & \cdots & \omega_{2N} \\
            \vdots & \vdots & \cdots & \vdots \\
            \omega_{N1} & \omega_{N2} & \cdots & \omega_{NN} \\
          \end{array}
        \right)
,~~
I_{\varepsilon}=\left(
                    \begin{array}{ccc}
                     I_{\epsilon_{1}} &   &   \\
                    & \ddots &   \\
                     &   & I_{\epsilon_{N}} \\
                                         \end{array}
                                       \right),
I_{\epsilon_{k}}=\left(
                        \begin{array}{cc}
                          I &   \\
                            & I \\
                        \end{array}
                      \right)
               _{2n_{k}\times2n_{k}},\\
(\omega_{j\ell})_{2n_{j}\times2n_{\ell}}&=\left(
                                           \begin{array}{cc}
                                             (\omega_{j\ell}^{(1)})_{n_{j}\times n_{\ell}} & (\omega_{j\ell}^{(2)})_{n_{j}\times n_{\ell}} \\
                                             (\omega_{j\ell}^{(3)})_{n_{j}\times n_{\ell}} & (\omega_{j\ell}^{(4)})_{n_{j}\times n_{\ell}} \\
                                           \end{array}
                                         \right),
(\widetilde{\omega_{j\ell}})_{2n_{j}\times2n_{\ell}}=\left(
                                           \begin{array}{cc}
                                             (\omega_{j\ell}^{(5)})_{n_{j}\times n_{\ell}} & (\omega_{j\ell}^{(6)})_{n_{j}\times n_{\ell}} \\
                                             (\omega_{j\ell}^{(7)})_{n_{j}\times n_{\ell}} & (\omega_{j\ell}^{(8)})_{n_{j}\times n_{\ell}} \\
                                           \end{array}
                                         \right),
\end{align}\end{subequations}
\begin{thm}
The multiple solitons of the focusing nonlocal mKdV equation with the NZBCs is derived by
\begin{align}
q(x,t)=q_{-}-i\left[\frac{\det \left((I_{\epsilon}-\widetilde{\Xi}\Xi)+\Gamma_{2}\langle W_{0}|\right)+\det\left((I_{\epsilon}-\widetilde{\Xi}\Xi)+\widetilde{\Xi}\Gamma_{1}\langle W_{0}|\right)} {\det(I_{\epsilon}-\widetilde{\Xi}\Xi)}-2\right],
\end{align}
where
\begin{align*}
\langle W_{0}|=\left(\langle W_{0}^{1}, \langle W_{0}^{2}|,\cdots,\langle W_{0}^{N}\right)^{T},
\langle W_{0}^{k}|=\left(1,0,\cdots,0,1,0,\cdots,0\right)_{1\times2n_{k}}.
\end{align*}
\end{thm}

\section{Conclusion}
In summary, we have studied the soliton solutions of nonlocal equations \eqref{Eq1} with two different boundary conditions and solved the initial problem. The scattering problem have analyzed by introducing uniformization variables. In the direct scattering part, the analyticity and symmetry properties of the Jost solution and scattering data, as well as the discrete spectrum are derived. The inverse scattering problem have been solved by the Riemann-Hilbert problem. Since the Plemelj formula have been used to solve the solution of the regular RH problem, our work is mainly to solve the irregular RH problem. Through the Laurent expansion of the RH problem, the soliton solution formula have established to solve the situation that the transmission coefficients have one high-order pole and multiple high-order poles, respectively. Finally, we consider the propagation behavior of one-soliton solution with $\delta\sigma=1$ and two-soliton solution with $\delta=\sigma=-1$. This is because the discrete spectrum have appeared in pairs with non-zero boundary conditions. In this case, the paired discrete spectrum have regarded as two different discrete spectra with the same power. For $\delta\sigma=1$, the discrete spectrum is the set $\left\{z_n, -z_n^*, \frac{q_{0}^2}{z_n}, -\frac{q_{0}^2}{z_n^*}\right\}$ while for $\delta\sigma=-1$ the discrete spectrum is the set $\left\{z_n, -z_n^*, -\frac{q_{0}^2}{z_n}, \frac{q_{0}^2}{z_n^*}\right\}$, we can summarize the discrete spectrum in these two different cases as a characteristic of discrete spectrum, which is $\left\{v_1, -v_1^*, v_2, -v_2^*\right\}$. In \cite{PD-2019}, we know that when $\sigma=1, \delta=-1$, there is no reflectionless potential of  the defocusing nonlocal mKdV equation.
Consequently, we only consider the above situation.

It is complex to deduce the soliton solution with $N$ high-order poles by the method in \cite{JMP-2014}. In our work, we mainly have adopted the following ideas. We have combined with the asymptotic properties of Jost function and scattering data, the piecewise meromorphic function $M(z)$ have been expanded at discrete spectral points, and have combined with the relationship between the solution of nonlocal equation \eqref{Eq1} and the solution of RH problem. Once the coefficients of piecewise meromorphic function $M(z)$ have been solved, the exact expression of the solution of the original equation have been obtained. In what follows, we expand the inverse scattering coefficient $r(k)$ of the pole by Laurent series and the oscillation term in the RH problem by Taylor expansion. Finally, a closed algebraic system is obtained. Through Cramer's law,  the exact expression of the solution for the original equation in the case of two reflection coefficients have been derived.
\section*{Acknowledgments}
This work was supported by the National Natural Science Foundation of China under Grant No. 11975306, the Natural Science Foundation of Jiangsu Province under Grant No. BK20181351, the Six Talent Peaks Project in Jiangsu Province under Grant No. JY-059,  the Fundamental Research Fund for the Central Universities under the Grant Nos. 2019ZDPY07 and 2019QNA35, and the Postgraduate Research \& Practice Innovation Program of Jiangsu Province under Grant No. KYCX21\_2152.

\end{document}